\documentclass[12pt,a4paper]{article}
\usepackage{jcappub}
\usepackage{amsmath}
\usepackage{amssymb}
\usepackage{epsfig}
\usepackage{latexsym,amsmath,amssymb}
\usepackage{graphicx}
\usepackage{float}
\usepackage{slashed}
\usepackage{bm}
\usepackage{epsfig}
\usepackage{dcolumn}
\usepackage{color}
\usepackage{subfigure}

\def\beq{\begin{equation}}
\def\eeq{\end{equation}}
\def\ber{\begin{eqnarray}}
\def\eer{\end{eqnarray}}
\def\benu{\begin{enumerate}}
\def\eenu{\end{enumerate}}
\def\vphi{\varphi}

\def\n{\nabla}

\def\mp{m_{p}}

\def \lleq {\lower0.9ex\hbox{ $\buildrel < \over \sim$} ~}
\def \ggeq {\lower0.9ex\hbox{ $\buildrel > \over \sim$} ~}

\def\asta{{Astron.\@ Astrophys.\ }}

\def\prl{{Phys.\@ Rev.\@ Lett.\ }}
\def\prd{{Phys.\@ Rev.\@ D\ }}

\def\plb {{Phys.\@ Lett.\@ B\ }}

\def\ie {{\it ie}}
\def\n {\noindent}

\def\om0{\Omega_{0m}}

\def \wp {w_{\varphi}}

\def \rcr {\rho_{cr,0}}

\def \lcdm {$\Lambda$CDM}

\title{New tracker models of dark energy}
\author[a]{Satadru Bag,}
\author[a]{Swagat S. Mishra}
\author[a]{and Varun Sahni}

\affiliation[a]{Inter-University Centre for Astronomy and Astrophysics,
Post Bag 4, Ganeshkhind, Pune 411~007, India}

\emailAdd{satadru@iucaa.in}
\emailAdd{swagat@iucaa.in}
\emailAdd{varun@iucaa.in}

\date{\today}

\abstract{
We describe a new class of dark energy (DE) models which behave like cosmological trackers at early times.
These models are based on the $\alpha$-attractor set of potentials, originally discussed
in the context of inflation. The new models allow the current acceleration of the universe to
be reached from a wide class of initial conditions.
Prominent examples of this class of models are the potentials
$\coth\vphi$ and  $\cosh\vphi$. 
A remarkable feature of this new class of models is that they lead to 
large enough negative values of the equation of state at the present epoch, consistent with the observations of accelerated expansion of the universe,  from a 
very large initial basin of attraction. They therefore avoid the fine tuning problem
which afflicts many models of DE. }


\begin{document}
\maketitle

\section{Introduction}
\label{sec:intro}

A remarkable property of our universe is that it appears to be accelerating.
Within the context of
Einstein's theory of general relativity, cosmic acceleration can arise if at least one of the
constituents of the universe violates the strong energy condition $\rho + 3p \geq 0$.
Physical models with this property are frequently referred to as `dark energy' (DE).
Although several models of DE have been advanced in the literature, 
perhaps the simplest remains Einstein's original idea of the cosmological constant,
$\Lambda$. As its name suggests, the energy density associated with the cosmological constant,
$\frac{\Lambda}{8\pi G}$, and its
 equation of state, $w = -1$, remain the same at all cosmological epochs.
Although $w = -1$ satisfies current observations very well, the non-evolving nature of
$\Lambda$ implies an enormous difference in its density and that in matter or
 radiation at early times. 
For instance $\rho_\Lambda/\rho_r \sim 10^{-58}$ at the time of the
electroweak phase transition, at earlier times this ratio is still smaller.

This `imbalance' between the non-evolving and small value of $\Lambda$ on the one hand, and the evolving 
density in matter/radiation on the other,
has fueled interest in models in which,
like matter/radiation, DE also evolves with time \cite{ss00,DE1,DE}.
In this context, considerable attention  has been focused on models with `tracker' properties
which enable the present value of the DE density to be reached from a wide range of initial conditions. This class of models appears to
alleviate the so-called `fine-tuning' (or `initial value') problem which characterizes
$\Lambda$ \cite{ratra,wetterich88,ferreira,zlatev,zlatev1,sw00,brax,barreiro,albrecht00}.
A scalar field with the inverse power-law (IPL) potential 
$ V \propto \vphi^{-\alpha}$ ($\alpha > 0$),
 presents one of the oldest and best
studied examples of this class of models \cite{ratra,wetterich88}.
Unfortunately the IPL model 
cannot account for the large negative values of $w_{\rm DE}$ at the present epoch consistent with the
observations \cite{BAO,SDSS,planck,asen17,huterer17} while at the same time 
preserving a large initial basin of attraction \cite{zlatev,zlatev1}.

In this paper we describe a new class of DE models based on the $\alpha$-attractors \cite{linde1,linde2}.
A compelling feature of these new  models is that they have a
very wide basin of attraction which allows the late time asymptote $w =-1$
to be reached from a large class of initial conditions.
A total of four different DE models are described in this paper.
Each of these models has very distinctive features which are reflected in the
evolution of $w_\vphi(z)$
and its first derivative, $w' = dw_\vphi/d lna$.
An interesting property of these models is that their current 
equation of state (EOS) 
can drop below $-0.9$, providing good agreement with present observational bounds.
Our results lead us to conclude that tracker models of DE could be very relevant for 
the understanding of cosmic acceleration. 

The plan of our paper is as follows:
The $\alpha$-attractor family of potentials 
is briefly discussed in section \ref{sec:alphainf}.
Section \ref{sec:DEmodels}
contains our main results and provides an analysis of the four new models of
tracker dark energy.  
A summary of our results is presented in
section \ref{sec:Summary}.

\section{Conformal inflation and $\alpha$-attractors}\label{sec:alphainf}

Kallosh \& Linde recently discovered an interesting new family
of potentials which could give rise to successful inflation \cite{linde1}. 
They noted 
that the Lagrangian
\beq
{\cal L} = \sqrt{-g}\left[ \frac{1}{2}\partial_\mu\chi\partial^\mu\chi +
\frac{\chi^2}{12}R(g) - \frac{1}{2}\partial_\mu\phi\partial^\mu\phi
-\frac{\phi^2}{12}R(g) - \frac{\tilde\lambda }{4}\left (\phi^2 - \chi^2\right )^2\right]
\, , \label{eq:1}
\eeq
where $\tilde\lambda$ is a dimensionless parameter and $\chi$, $\phi$ are scalar fields, 
is invariant under the ${\rm O}(1, 1)$ group of transformations in the $(\chi,
\phi)$ space and also under the group of local conformal transformations. Fixing the local
conformal gauge to
\beq \label{eq:gauge-fix}
\chi^2-\phi^2 = 6 m_{p}^2 \, ,
\eeq
the Lagrangian in (\ref{eq:1}) can be parameterized by
\beq
\chi = \sqrt{6} m_p \cosh{\frac{\vphi}{\sqrt{6} m_p}}~, \quad \phi = \sqrt{6} m_p
\sinh{\frac{\vphi}{\sqrt{6} m_p}} \, ,
\eeq
so that
\beq
\frac{\phi}{\chi} = \tanh{\frac{\vphi}{\sqrt{6} m_p}}~.
\eeq
Consequently (\ref{eq:1}) reduces to
\beq
{\cal L} = \sqrt{-g}\left\lbrack \frac{m_{p}^2}{2} R -
\frac{1}{2}\partial_\mu\vphi\partial^\mu\vphi - \Lambda  m_{p}^2 \right\rbrack \,
, \label{eq:deS}
\eeq
which describes general relativity with
the cosmological constant $\Lambda = 9\tilde\lambda m_{p}^2$
 (here $m_p = 1/ \sqrt{8 \pi G}
\approx 2.4 \times 10^{18}$~GeV). 
A conformally invariant generalization of (\ref{eq:1}) and (\ref{eq:deS}), $\Lambda  \to
F(\phi/\chi)m_p^2$, with an arbitrary function $F$ that deforms the ${\rm O} (1, 1)$
symmetry, results in the Lagrangian
\beq
{\cal L} = \sqrt{-g}\left\lbrack \frac{m_{p}^2}{2} R -
\frac{1}{2}\partial_\mu\vphi\partial^\mu\vphi - V (\varphi) \right\rbrack
\label{eq:L2}
\eeq
with the scalar-field potential
\beq \label{eq:pot}
V (\varphi) = m_{p}^4 F \left( \tanh{\frac{\vphi}{\sqrt{6} m_p}} \right) \, .
\eeq
Different canonical potentials $V(\vphi)$ were discussed in \cite{linde1} in the
context of inflation, while \cite{linde2} introduced the $\alpha$-attractor
 family of potentials following
the prescription\footnote{The parameter $\alpha$ can
be related to the curvature of the superconformal K$\ddot{\rm a}$hler metric
\cite{linde2}.} $V (\varphi) \to V \left( \varphi /
\sqrt{\alpha} \right)$.
An attractive feature of the $\alpha$-attractors is that they
are able to parameterize a wide variety of inflationary scenarios within a common setting.

In \cite{sss17} it was shown that, in addition to defining inflationary models, the
$\alpha$-attractors were also able to source oscillatory models of dark matter and dark energy\footnote{DE 
from $\alpha$-attractors has also been discussed in \cite{linder_alpha}, although not in
the tracker context. }.
In this paper we extend the study of \cite{sss17} by showing that
the $\alpha$-attractors can give rise to new models of tracker-DE in which 
the equation of state (EOS) approaches the
late-time value $w \simeq -1$ from a wide class of initial conditions.


\n
In this paper our focus will be on the $\alpha$-attractor family of potentials
characterized by 
\beq \label{eq:pot1}
V (\varphi) = m_{p}^4 F \left( \tanh{\frac{\vphi}{\sqrt{6\alpha} m_p}} \right) \, .
\eeq

Accordingly, our dark energy models are based on the following potentials all of which have
interesting tracker properties:

\begin{figure}[htb]
\centering
\subfigure[][]{
\includegraphics[width=0.485\textwidth]
{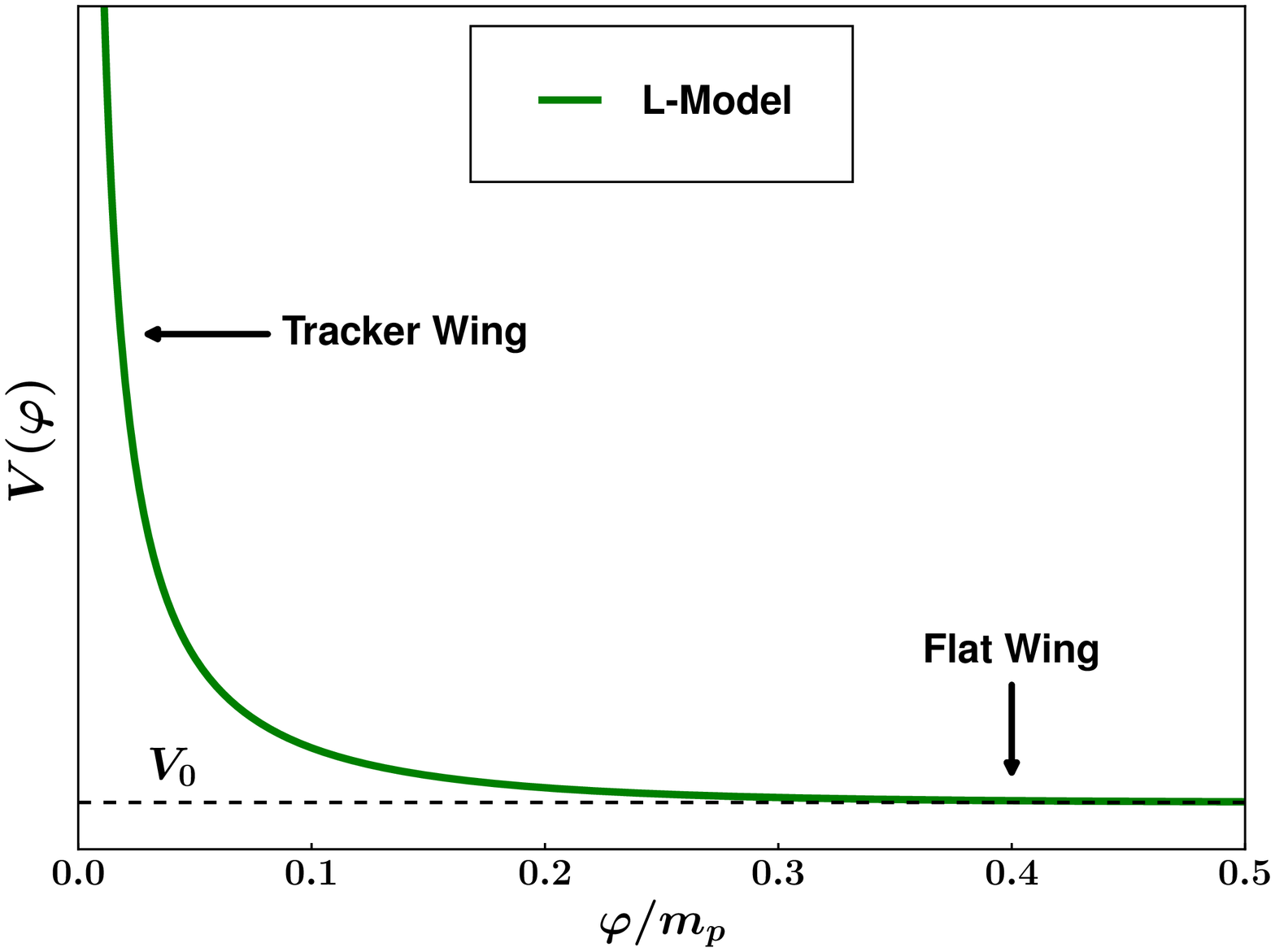}\label{fig:DE_potcoth}}
\subfigure[][]{
\includegraphics[width=0.485\textwidth]
{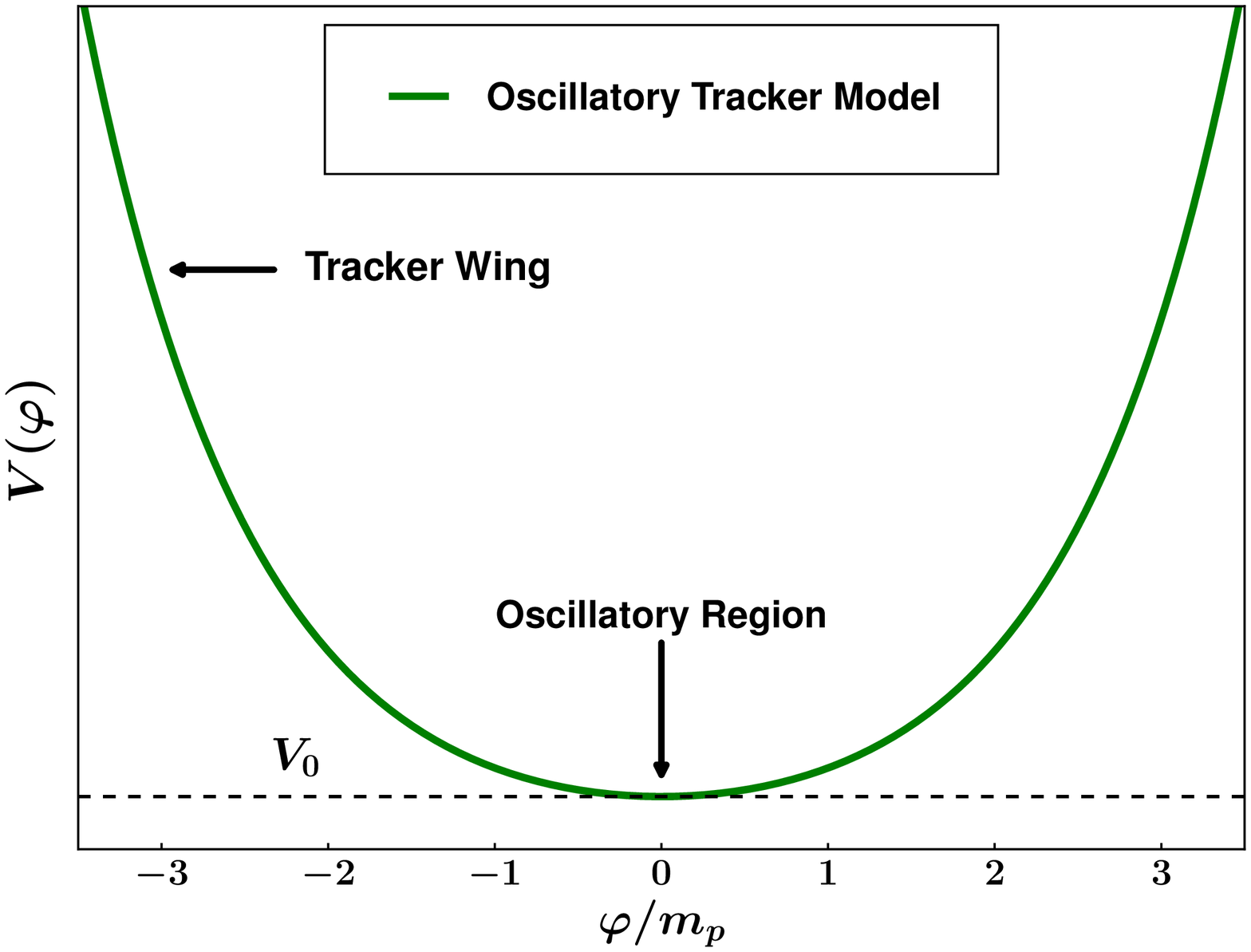}\label{fig:DE_potcosh}}
\subfigure[][]{
\includegraphics[width=0.485\textwidth]
{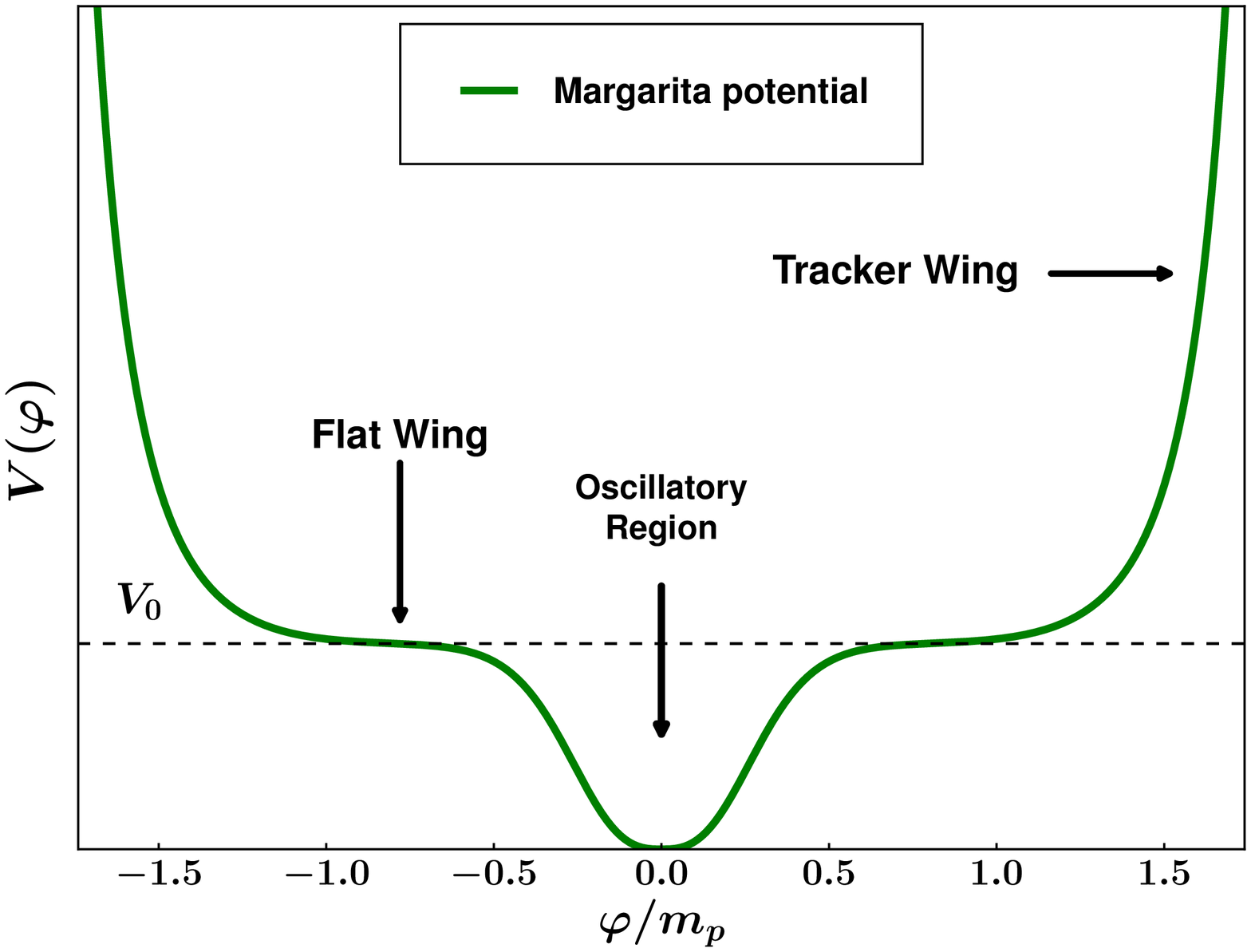}\label{fig:DE_potchamp}}
\subfigure[][]{
\includegraphics[width=0.485\textwidth]
{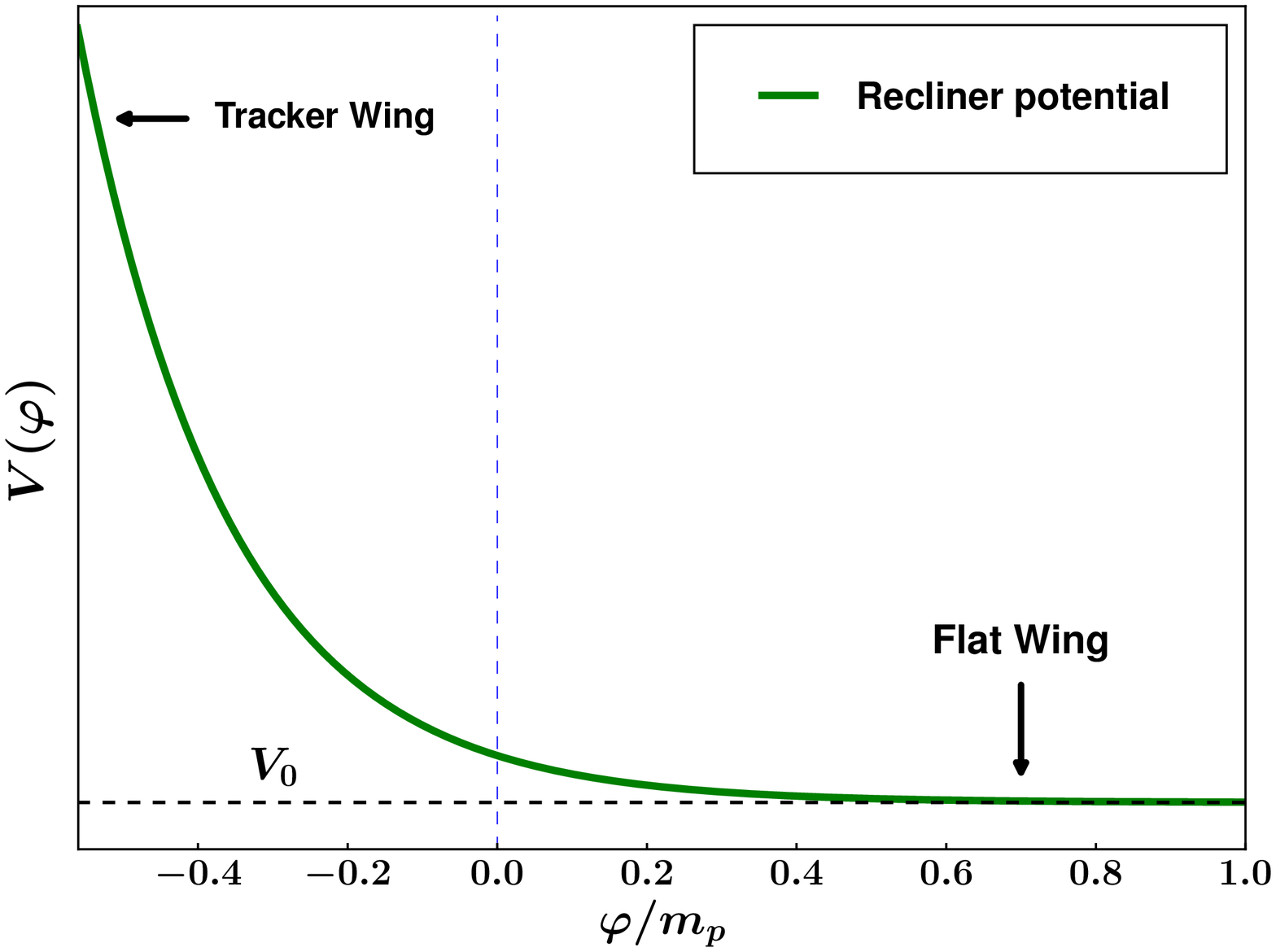}\label{fig:DE_potexp}}
\caption{Potentials corresponding to different tracker models of dark energy
are schematically displayed.  Clockwise from the upper left:
The L-model (\ref{eq:PLT}), the Oscillatory tracker model (\ref{eq:OLT}),
the Recliner potential (\ref{eq:exp})
and
the Margarita potential (\ref{eq:champagne}).
} \label{fig:potentials}
\end{figure}

\begin{enumerate}

\item The L-model

\beq
V(\vphi) = V_0\, \coth{\left(\frac{\lambda\vphi}{m_p}\right)} 
\label{eq:PLT}
\eeq
equivalently $V(\vphi) = V_0\,\left\lbrack\tanh{\frac{\lambda\vphi}{m_p}}\right\rbrack^{-1}$.

\item The Oscillatory tracker model

\beq
 V(\vphi) = V_0\, \cosh{\left(\frac{\lambda\vphi}{m_p}\right)} 
\label{eq:OLT}
\eeq
equivalently $V(\vphi) = V_0\,\left[ 1-\tanh^2\left(\frac{\lambda\vphi}{m_p}\right) \right]^{-1/2}$.

\item The Recliner model

\beq
V(\vphi) = V_0\, \left \lbrack 1+ \exp{(-\frac{\lambda\vphi}{m_p})}\right \rbrack 
\label{eq:exp}
\eeq
equivalently $V(\vphi) = V_0\,\left[ 1+ \tanh{\left(\frac{\lambda\vphi}{2m_p}\right)}\right] ^{-1}$.

\item The Margarita potential

\beq
V(\vphi) = V_0 \tanh^2{\left(\frac{\lambda_1\vphi}{m_p}\right)}\cosh{\left(\frac{\lambda_2\vphi}{m_p}\right)} 
\label{eq:champagne}
\eeq
where $\lambda_1\gg\lambda_2$.

\end{enumerate}

\n
The tracker parameter $\lambda$ in our models is related to the parameter $\alpha$ in
(\ref{eq:pot1}) by
\beq
\lambda = \sqrt{\frac{1}{6\alpha}}~.
\eeq

\n
The reader might like to note that in terms of the variable 
$x = \tanh{\vphi}$, the original $\alpha$-attractor based T-model of Inflation
\cite{linde1} is simply
$F(x) = x$ in \eqref{eq:pot1}, while our L-model is $F(x) = 1/x$.
The functional form of $F(x)$ for
our remaining three DE models is somewhat
more complicated: $F(x) = (1-x^2)^{-1/2}$, for the oscillatory tracker \eqref{eq:OLT};
$F(x) = \frac{1}{1+x}$ for the recliner potential \eqref{eq:exp}; and finally
$F(x) = \frac{x^2}{\sqrt{1-x^2}}$ for the Margarita potential \eqref{eq:champagne}.
Therefore, from the $\alpha$-attractor perspective, the L-model \eqref{eq:PLT} appears to be
the most appealing of the four DE models introduced by us.

\n
One should also point out that, in addition to the above four models, hyperbolic potentials
have also been used in connection with the following models of dark energy.

\begin{itemize}

\item  DE with a constant equation of state $-1<w<0$ is described by
\cite{ss00,DE1}

\beq
V(\vphi)={3H_0^2(1-w)(1-\Omega_{m0})^{1/|w|} \over 16\pi G\Omega_{m0}^\alpha}
\left\lbrack\sinh\left(|w|\sqrt{{6\pi G\over 1+w}}
(\vphi-\vphi_0+\vphi_1)\right)\right\rbrack^{-2\alpha}~,
\label{quequi}
\eeq
where
\beq
\alpha = \frac{1+w}{|w|},~~\vphi_0=\vphi(t_0),~~\vphi_1= \sqrt{{1+w\over
6\pi G}}{1\over |w|} \ln {1+\sqrt{1-\Omega_{m0}}\over
\sqrt{\Omega_{m0}}}~. \nonumber
\eeq

\item The Chaplygin gas with
$p = - A/\rho$ can be described by the scalar field potential
\cite{chap,chap1} 
\beq
V(\vphi) = \frac{\sqrt{A}}{2}\left (\cosh(2\sqrt{6\pi G}\vphi) + \frac{1}
{\cosh(2\sqrt{6\pi G}\vphi)}\right )~.
\eeq
Note that the 
Chaplygin gas can also be modelled using a scalar field
with the Born-Infeld kinetic term \cite{bilic,frolov}. 

\end{itemize}

It is interesting that all four of the dark energy models introduced  in our paper
possess distinct features which allow them to be distinguished 
from each other at late times.
We shall elaborate on these models in the next section.

\section{Tracker Models of Dark Energy}
\label{sec:DEmodels}

\subsection{L-model}
\label{sec:Inv_power_trac}
 
Consider first the L-model (\ref{eq:PLT})
and its natural extension\footnote{We refer to this model as the `L-model'
since $V(\varphi)$ has a characteristic L shape for large values of $\lambda$ and $p$
as shown in figure \ref{fig:potentials}.}
\beq
V(\vphi) = V_0\, \coth^p{\left(\frac{\lambda\vphi}{m_p}\right)}~.
\label{eq:coth}
\eeq
For small values of the argument, $0 < \frac{\lambda\vphi}{m_p} \ll 1$, one finds
\beq
V \simeq \frac{V_0}{(\lambda\vphi/m_p)^p}
\label{eq:IPL}
\eeq
which suggests that the early time behaviour of this model is very similar
to that of the IPL model for which, at early times \cite{ratra,zlatev}
\beq \label{eq:wtrack_ipl}
w_\vphi = \frac{pw_{\rm B} - 2}{p+2}~,
\eeq
where $w_{\rm B}$ is the background equation of state of matter/radiation.
The IPL model \eqref{eq:IPL} therefore has the appealing property that,
for large values of $p \gg 1$,
its EOS can track the EOS of the dominant matter component in the universe.
Unfortunately it is also well known that, for $\Omega_{0m} \geq 0.2$, the IPL model 
(\ref{eq:IPL}) with $p > 1$
cannot give rise to $w_0 < -0.8$ at the present epoch \cite{zlatev1}.
This may be viewed as a significant shortfall of this model since
observations appear to suggest that the current EOS of dark energy should
satisfy $w_0 \leq -0.9$ \cite{BAO,SDSS,planck,asen17,huterer17}.
Of course this problem can be bypassed if one assumes a smaller value
$p < 1$ for the exponent in (\ref{eq:IPL}). However in this case
the initial basin of attraction shrinks considerably, which diminishes the
appeal of the IPL model. 

\n
In contrast to the IPL model, the L-potential
has the following asymptote for
$\frac{\lambda\vphi}{m_p} \gg 1$
\beq
V(\varphi)
\simeq V_0
\label{eq:coth_asymp}
\eeq
indicating that the L-potential flattens and
begins to behave like a cosmological constant
at late times. Because of this the present value of the EOS
in the L-model can be significantly lower\footnote{Another means
of lowering $w_0$ is by coupling $\vphi$ to the Ricci scalar, as shown in \cite{matarrese}.} than that in the IPL model (\ref{eq:IPL}).

\begin{figure}[htb]
\centering
\includegraphics[width=0.85\textwidth]{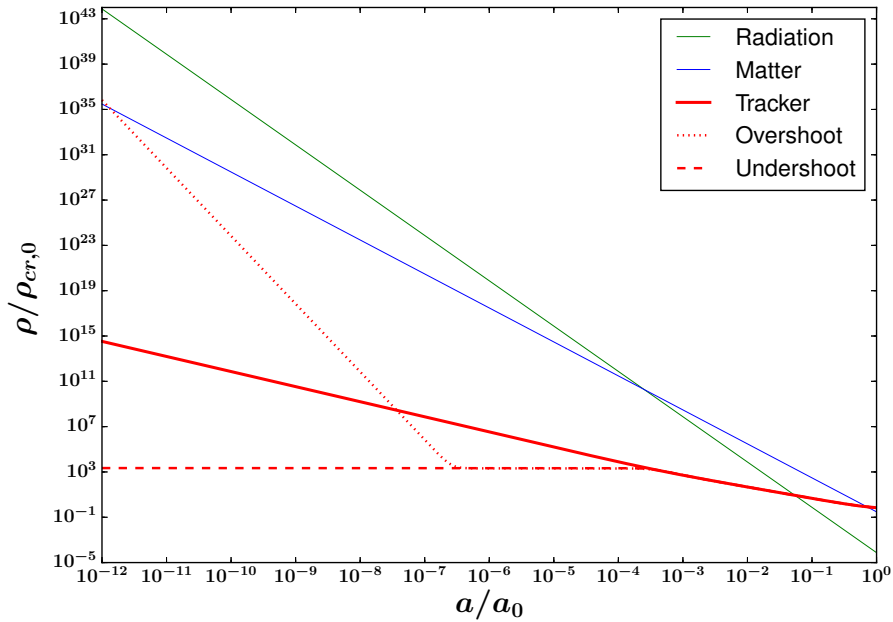}
\caption{This figure illustrates the tracking feature of the L-model
 $V=V_0 \coth (\varphi/\mp)$. The density of the different components is
plotted against the scale factor; here $\rcr \equiv 3 \mp^2 H_0 ^2$ is the current
 critical density. The solid red curve represents the tracker branch which is 
similar to that of the corresponding  IPL model at early times. 
Trajectories associated with ``overshoot'' (dotted red) and ``undershoot'' 
(dashed red) initial conditions meet the tracker branch by
$z \sim 10^3$. }
\label{fig:coth1_l1p1_tracker_Ophi}
\end{figure}

\begin{figure}[htb]
\centering
\subfigure[]{
\includegraphics[width=0.47\textwidth]{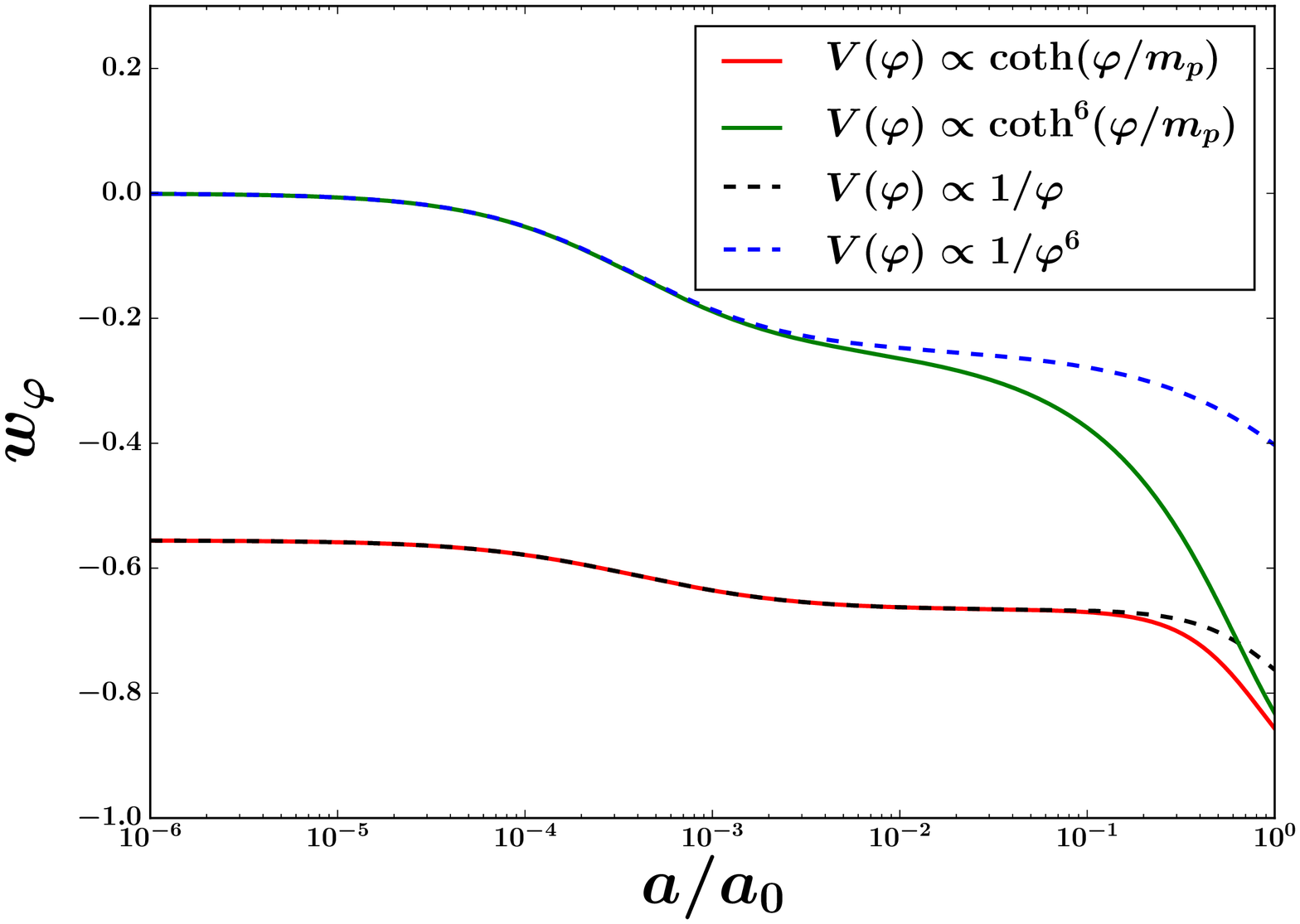}\label{fig:w_ipl_coth1}}
\subfigure[]{
\includegraphics[width=0.47\textwidth]{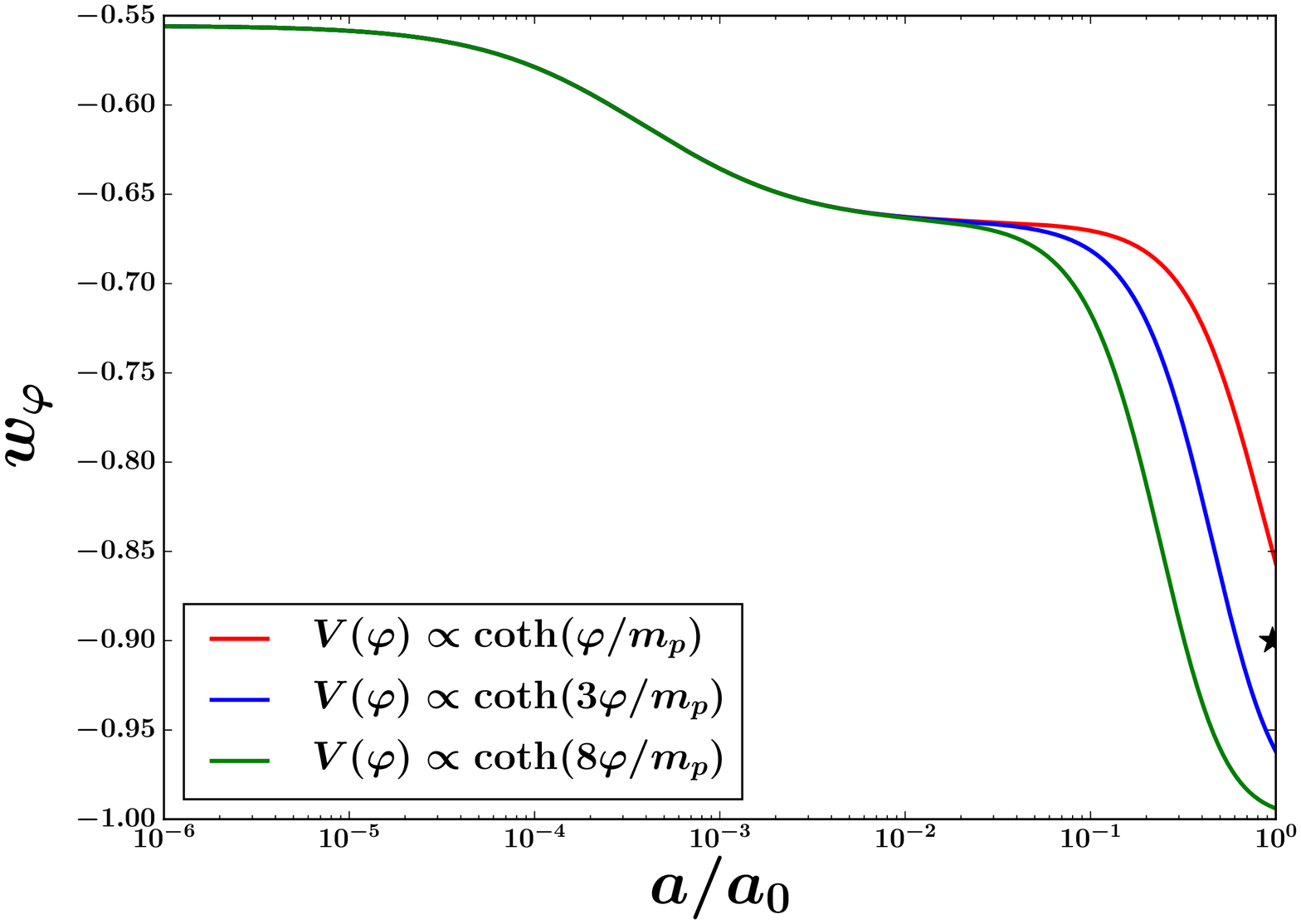}\label{fig:w_ipl_coth1_p1_lam}}
\caption{{\bf (a):} This figure illustrates that the
 L-model \eqref{eq:coth} possesses early time tracking features 
similar to that of the
 inverse power law potential \eqref{eq:IPL}. 
However at late times, $\wp$ in the L-model 
falls significantly below $\wp$ in the IPL model.
 Note that this figure is shown for illustrative purposes and does not reflect
current observational constraints on $\wp$, which are met by the models
in the right panel.
 {\bf (b):} Increasing
 the value of $\lambda$ in  \eqref{eq:coth} 
while keeping $p$ fixed,
makes the current equation of state drop
to more negative values.
The black star on the right y-axis indicates the (conservative) observational $2 \sigma$
upper bound on $w_0$ for DE with a slowly varying EOS \cite{huterer17}. 
}
\label{fig:coth1_lam_p_a}
\end{figure}

The behavior of the L-potential  \eqref{eq:coth} is illustrated in Fig. \ref{fig:coth1_l1p1_tracker_Ophi}. 
The evolution of the scalar field energy density has been
 determined by solving the following system of equations
relating to a spatially flat Friedman Robertson Walker (FRW) universe
\ber
H^2 = \frac{8\pi G}{3}\Bigl( \rho_m + \rho_r + \rho_\vphi \Bigr) 
\eer
\ber
{\ddot \vphi} + 3H{\dot\vphi} + V'(\vphi) = 0~,
\label{eq:eom}
\eer 
where $\rho_m$ ($\rho_r$) is the density of matter (radiation),
and the density and pressure of the scalar field are
\ber\label{eq:rho_p_phi}
\rho_\vphi = \frac{1}{2}{\dot\vphi}^2 + V(\vphi), \qquad p_\vphi =
\frac{1}{2}{\dot\vphi}^2 - V(\vphi)~.
\eer
  As expected, the early time tracking phase in our
 model \eqref{eq:coth} -- illustrated by figure \ref{fig:w_ipl_coth1}, is identical to 
the tracking phase of the corresponding IPL potential $V\propto 1/\vphi^p$.

Figure \ref{fig:w_ipl_coth1}
shows the evolution of $w_{\vphi}(z) = p_\vphi/\rho_\vphi$, in the
L-potential $V = V_0\coth^p{\vphi/m_p}$ and in the IPL potential 
$V= V_0\left(\frac{m_p}{\vphi}\right)^p$.
Note that both potentials have precisely two free parameters: $V_0$ and $p$.
Fig. \ref{fig:w_ipl_coth1} draws attention to the interesting fact that, for 
the L-model with
$V \sim \coth^6{\vphi}$, the current value of $w_\vphi$ can be as low as
$w_\vphi \sim -0.8$, which is considerably lower than the corresponding value 
$w_\vphi \sim -0.4$ for $V \sim \vphi^{-6}$.
In other words, for identical values of $p$,
the late-time value of $w_\vphi$ in the L-model (\ref{eq:coth}) is 
{\em significantly lower}
than that in the IPL model (\ref{eq:IPL}). 
The value of $w_\vphi$ can be further lowered by increasing the value of $\lambda$ in
(\ref{eq:coth}), as shown in  figure \ref{fig:w_ipl_coth1_p1_lam}.  The black star on the right y-axis of figure \ref{fig:w_ipl_coth1_p1_lam} indicates the observational $2\sigma$ upper bound,
$w_0 \leq -0.9$, for DE models with a slowly varying EOS\footnote{ The
current $2\sigma$ upper bound on the EOS varies between $-0.8$ to $-0.9$, depending upon the
data sets employed and the method of reconstruction
\cite{huterer17}. We assume the conservative bound $w_0 \leq -0.9$ to highlight
the fact that the EOS in our models can drop to sufficiently low values at late times.} \cite{huterer17}.

\begin{figure}[htb]
\centering
\subfigure[$V=V_0 \coth ^p ( \varphi/\mp)$]{
\includegraphics[width=0.47\textwidth]{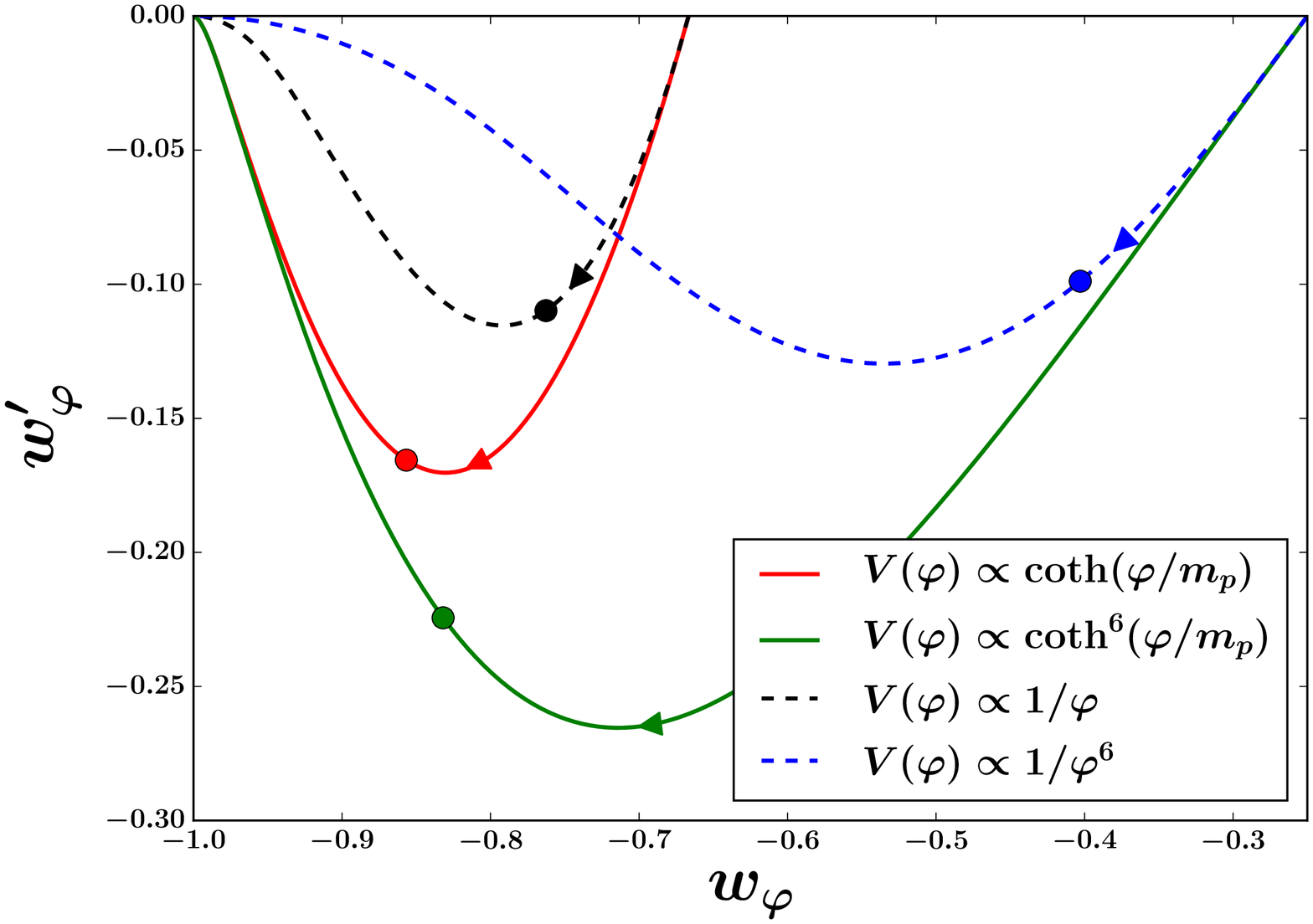}\label{fig:coth1_lam1_p}}
\subfigure[$V=V_0 \coth ( \lambda \varphi/\mp)$]{
\includegraphics[width=0.47\textwidth]{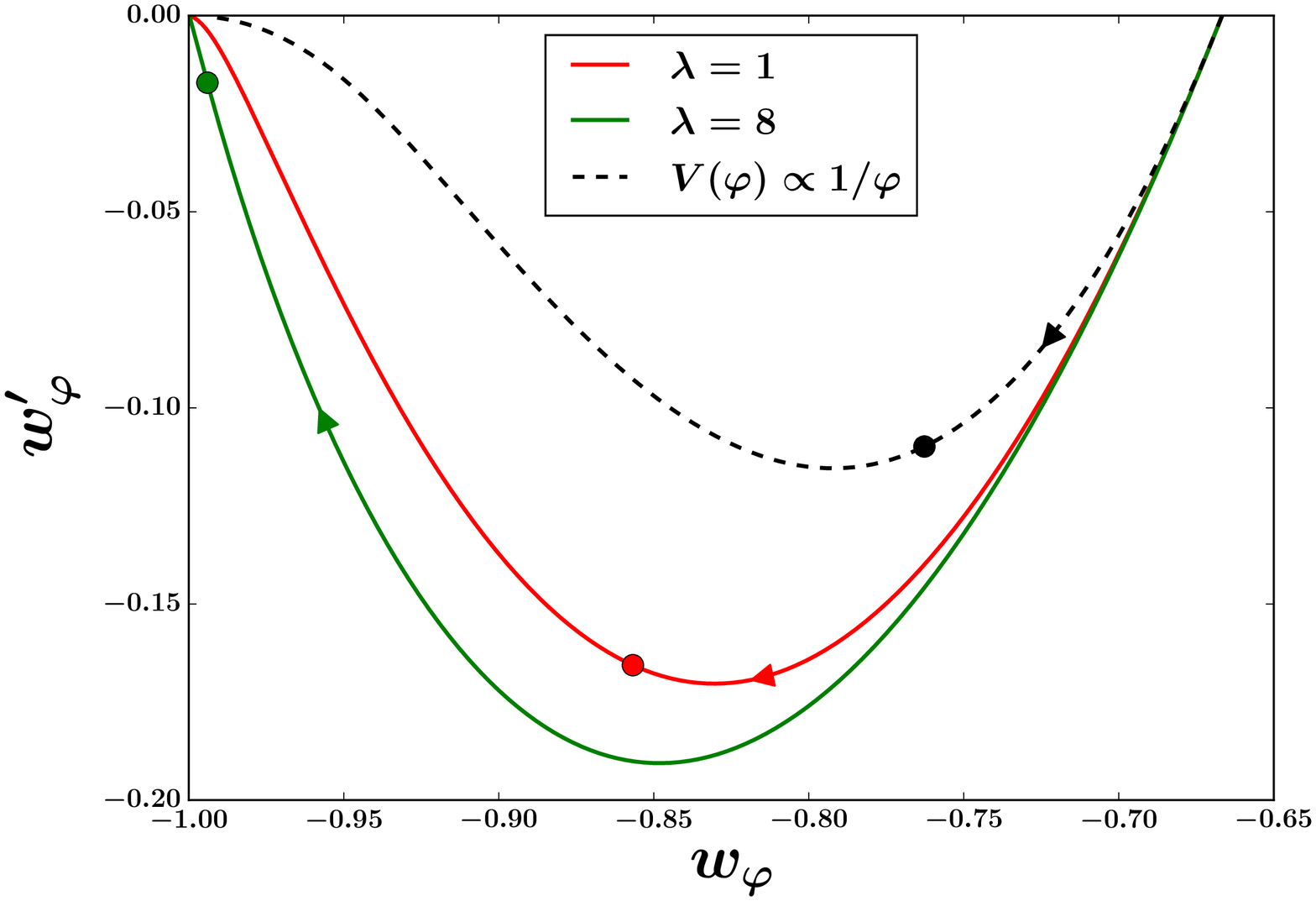}\label{fig:coth1_p1_lam}}
\caption{Phase space $\lbrace w_\varphi,w'_\varphi \rbrace$ 
trajectories for potentials \eqref{eq:coth} and \eqref{eq:IPL} starting from the matter dominated epoch.  The filled circle on each trajectory represents the present epoch with $\Omega_{0m}=0.3$. Both panels show that in the 
far future the EOS in all models approaches the \lcdm~ value of $w_\varphi=-1,w'_\varphi=0$. 
The {\bf left panel} demonstrates that with increasing values of $p$ 
the current value of
$w_\varphi$ increases while $w'_\varphi$ decreases. 
Note that, for identical values of $p$,
 the current values of $w_\vphi, w'_\varphi$ are much lower in L1 relative to
the IPL potential $V \propto \vphi^{-p}$.
From the {\bf right panel} one finds that an increase in the value of $\lambda$ leads to
a decrease in $w_\varphi$ and an increase in $w'_\varphi$. 
It is interesting that the different variants of the L-model \eqref{eq:coth}
can be easily distinguished from each other and from the IPL potential \eqref{eq:IPL}.
(Note that the left and right panels are not drawn to scale.)
}
\label{fig:coth1_lam_p}
\end{figure}

Figure \ref{fig:coth1_lam_p} shows the phase-space 
trajectories of the equation of state $\lbrace w_{\vphi},w_{\vphi}'\rbrace$ starting from the
matter dominated epoch.
Here \cite{caldwell_linder,linder17} 
\beq
w_{\vphi}' \equiv \frac{dw_{\vphi}}{d\ln a} = \frac{\dot w_{\vphi}}{H}.
\eeq
Note that all trajectories approach the \lcdm~ limit ($w_{\vphi}=-1,w_{\vphi}'=0$) at 
late times. The present epoch is marked by a circle on each trajectory.
Comparing the L-model \eqref{eq:coth} with the IPL potential \eqref{eq:IPL}
we find that the current EOS in the former is always more negative than that in the latter,
\ie~ $w_0^{\rm L1} < w_0^{\rm IPL}$, which supports our earlier results in 
fig. \ref{fig:w_ipl_coth1}.
Setting $\lambda = 1$ in \eqref{eq:coth} and increasing the value of $p$
leads to $w_{\vphi}'$ decreasing while $w_{\vphi}$ increases.
On the other hand, increasing $\lambda$ (with $p$ held fixed) leads to the opposite
behaviour, namely $w_{\vphi}$ decreases whereas $w_{\vphi}'$ increases.
Note that for moderately large values of $\lambda$ and $p$ 
($\lambda, p \sim {\rm few}$) the L-model \eqref{eq:coth} will have a large initial
basin of attraction before converging to $w_{\vphi}\sim -1,w_{\vphi}'\sim 0$ 
by the present epoch.

\subsection{Oscillatory tracker model}
\label{sec:Osc_trac}
 Next we turn our attention to the oscillatory tracker model (\ref{eq:OLT}), namely

\beq
 V(\vphi) = V_0\, \cosh{\left(\frac{\lambda\vphi}{m_p}\right)}~.
\label{eq:OLT1}
\eeq
For large values $\frac{\lambda|\vphi|}{m_p} \gg 1$, this potential has the asymptotic form
\beq
V \simeq \frac{V_0}{2} \exp{\left(\frac{\lambda\vphi}{m_p}\right)}.
\label{eq:exp0}
\eeq
The exponential potential has been extensively studied in \cite{ratra,ferreira,barreiro,sfs}.
In the context of a spatially flat FRW universe
it is well known that for $\lambda^2 > 3(1+w_{\rm B})$ the late time attractor in this model
has the same equation of state as the background fluid, namely 
$w_\vphi = w_{\rm B}$. The associated fractional density of the scalar field is
\beq
\Omega_\vphi = \frac{3(1+w_{\rm B})}{\lambda^2}~,
\eeq
with nucleosynthesis constraints imposing the lower bound $\lambda \ggeq 5$ \cite{ferreira,copeland98}, while the CMB constraints impose an even stronger lower bound $\lambda \geq 13$ \cite{CMB_Neff1,CMB_Neff2,scherrer_exp}.

\begin{figure}[htb]
\centering
\includegraphics[width=0.83\textwidth]{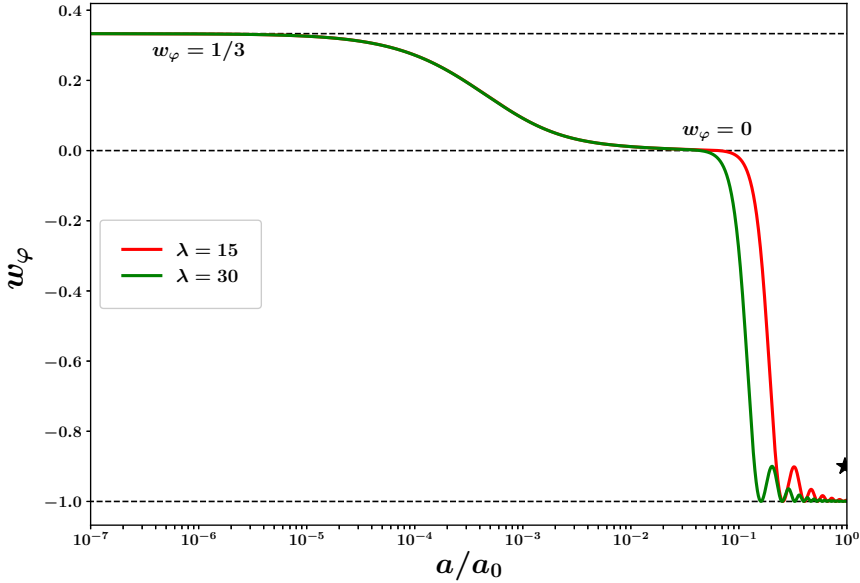}
\caption{This figure illustrates the evolution of the equation of state, $w_{\vphi}$,
 of the scalar field with the potential $V(\vphi) = V_0\, \cosh{\left(\frac{\lambda\vphi}{m_p}\right)}$, for two values of $\lambda$. 
During the radiation and matter dominated epochs, when 
$V \sim \exp{\left(\frac{\lambda\vphi}{m_p}\right)}$, the scalar field tracks the background density
resulting in $w_{\vphi}=1/3,~0$ respectively. At late times, after the commencement of
scalar field  oscillations, the EOS approaches $w_{\vphi}\simeq -1$, via small oscillations. 
 The black star on the right  indicates the (conservative) observational upper bound on $w_0$ (at $2 \sigma$) for DE models with a slowly varying EOS \cite{huterer17}.
We show this bound since the dynamical (oscillatory) contribution to $w_{\vphi}$
becomes quite small at low $z$ in our model.
}
\label{fig:DE_coshw}
\end{figure}

\begin{figure}[htb]
\centering
\subfigure[][]{
\includegraphics[width=0.485\textwidth]
{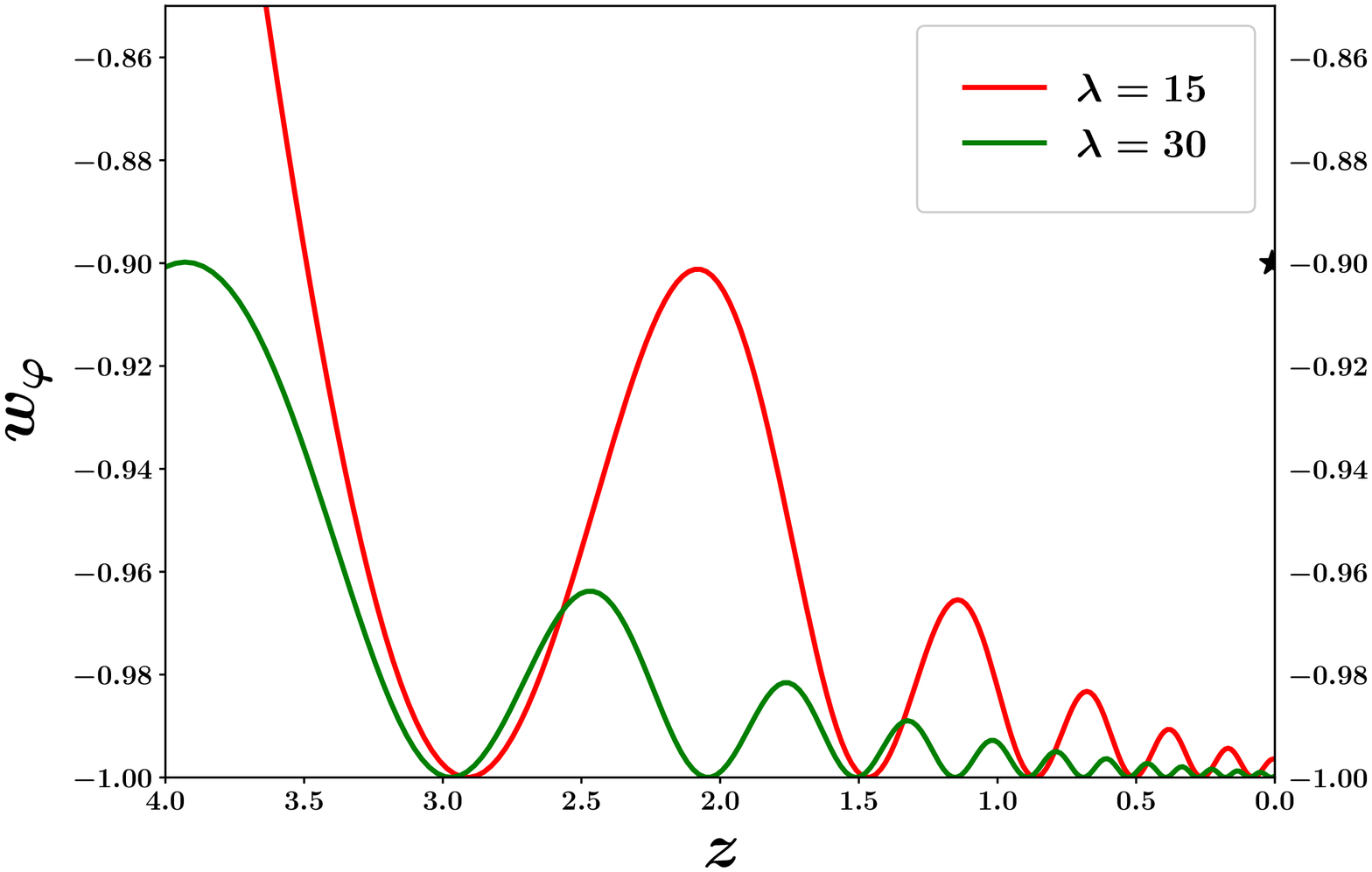}\label{fig:DE_coshRS}}
\subfigure[][]{
\includegraphics[width=0.485\textwidth]
{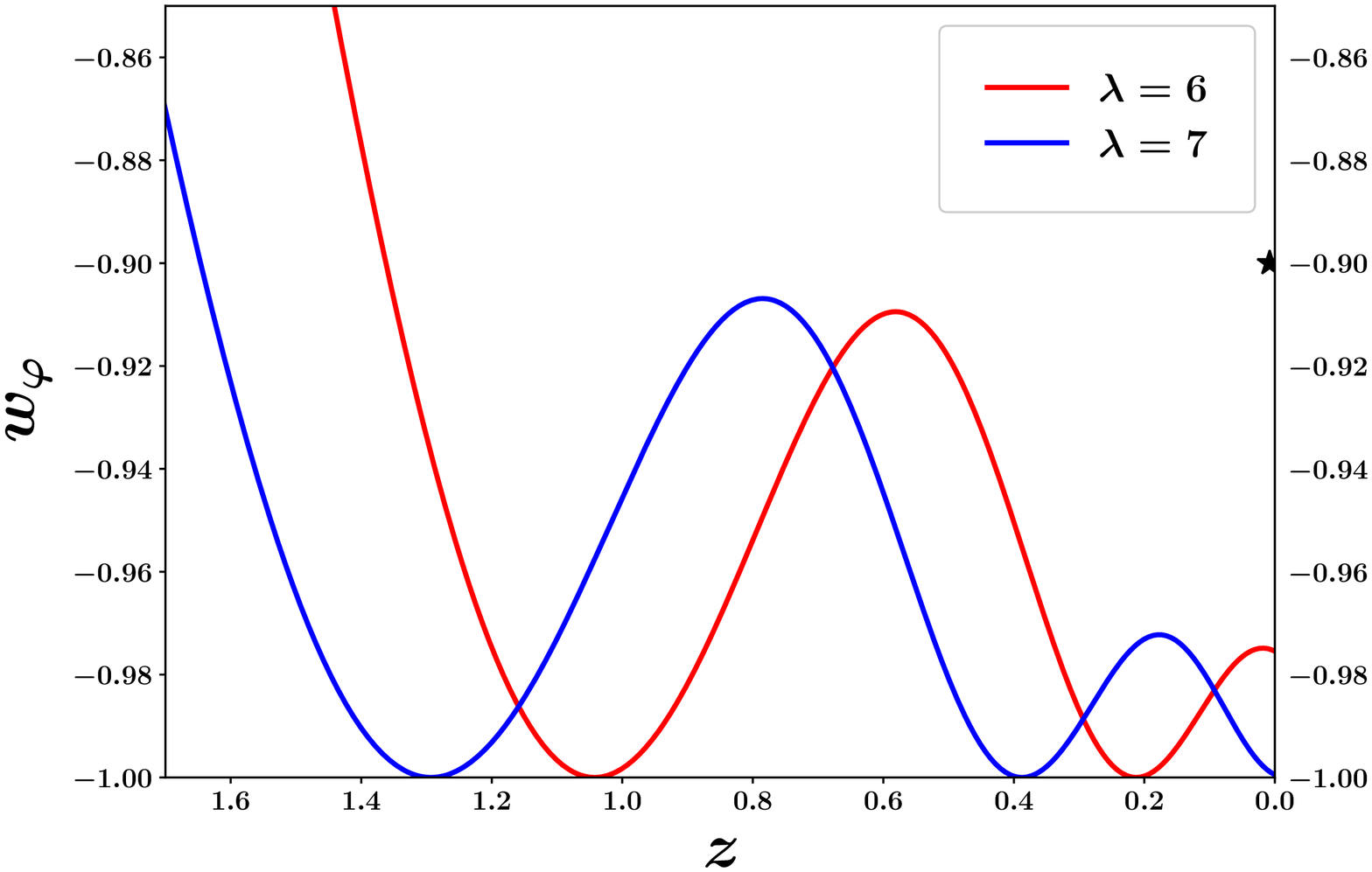}\label{fig:DE_coshRS1}}
\caption{{\bf Left panel:} The evolution of $w_{\vphi}(z)$ near
 the present epoch is shown 
for the potential $V(\vphi) = V_0\, \cosh{\left(\frac{\lambda\vphi}{m_p}\right)}$ 
with $\lambda=15$ and $20$.
One finds that although the behaviour of $w_{\vphi}(z)$ varies greatly between the two models,
all models have certain common features, for instance:
(i) the oscillation amplitude decreases with time;
(ii) the decreasing oscillation amplitude leads to
$w_{\vphi} \simeq -1$ at late times.
{\bf Right panel:} Models with $\lambda=6,7$ highlight the phase dependence
of oscillations for moderate values of $\lambda$.
 In both panels the black star on the right  indicates the (conservative) observational
upper bound on $w_0$ (at $2 \sigma$) for DE models with a slowly varying EOS \cite{huterer17}.}
 \label{fig:DE_coshwwz}
\end{figure}

\begin{figure}[htb]
\centering
\subfigure[][]{
\includegraphics[width=0.485\textwidth]
{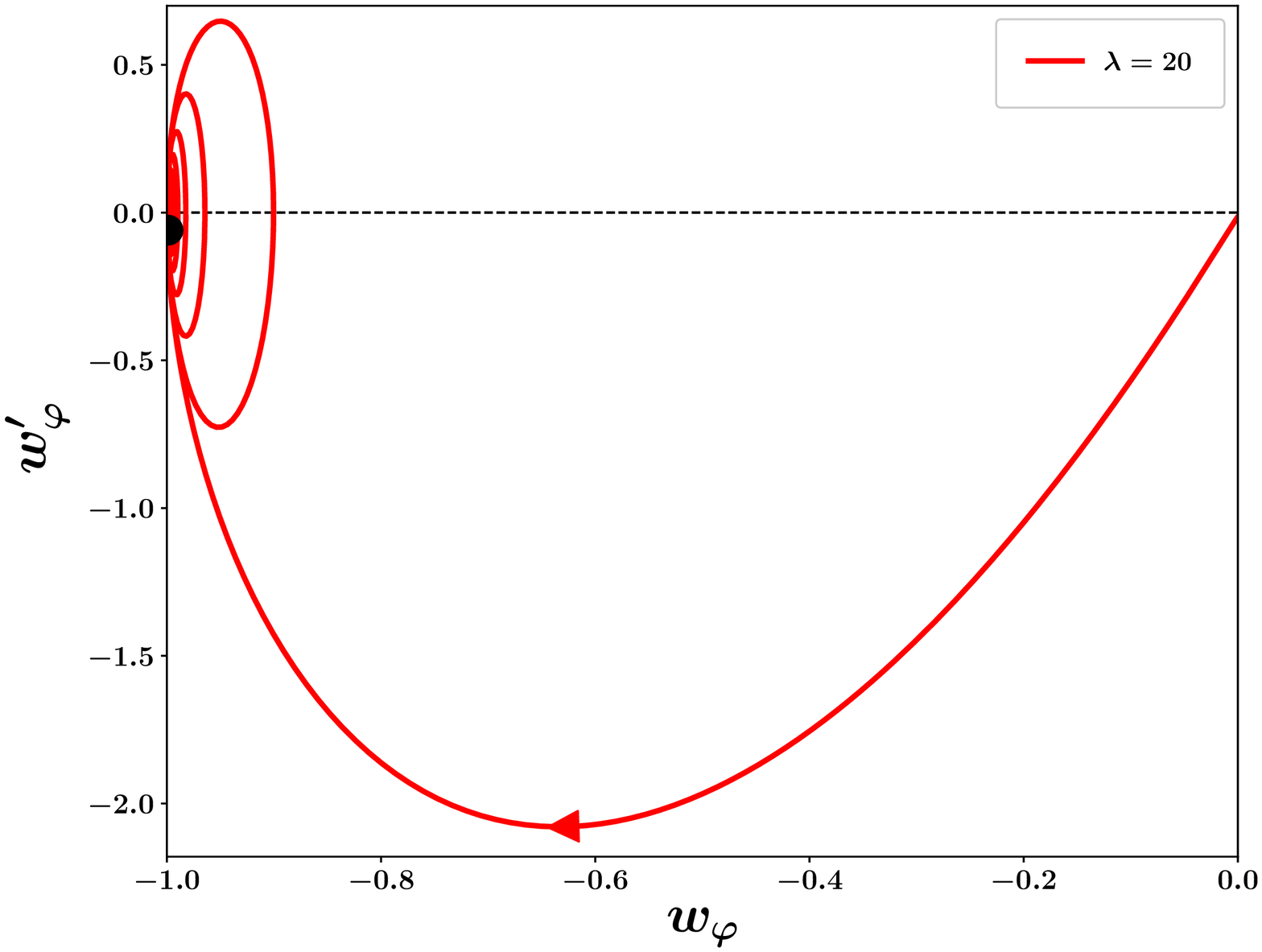}\label{fig:DE_coshww1}}
\subfigure[][]{
\includegraphics[width=0.485\textwidth]
{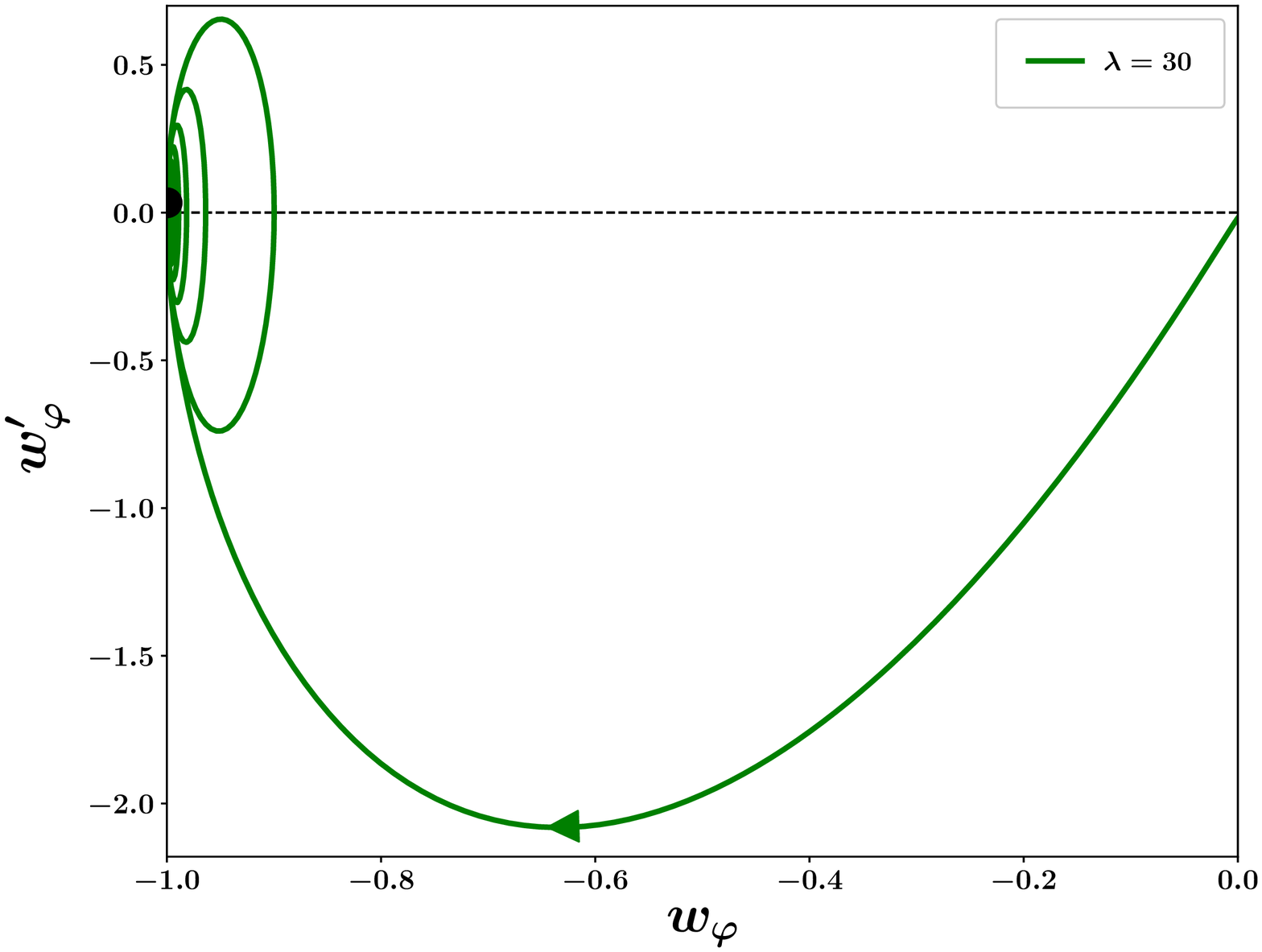}\label{fig:DE_coshww2}}
\caption{The phase-space plot of $\lbrace w_{\vphi},w_{\vphi}'\rbrace$ for the potential 
$V(\vphi) = V_0\, \cosh{\left(\frac{\lambda\vphi}{m_p}\right)}$ with $\lambda=20$
(left panel) and $\lambda = 30$ (right panel) starting from the matter dominated epoch. Arrows depict the flow of time
and black circles indicate the present epoch
with $\Omega_{0m} = 0.3$.
} \label{fig:DE_coshww}
\end{figure}

For small values, $\frac{\lambda|\vphi|}{m_p} \ll 1$, the potential (\ref{eq:OLT1}) has 
the limiting form
\beq 
V(\vphi) \simeq V_0\left\lbrack 1 + \frac{1}{2}\left(\frac{\lambda\vphi}{m_p}\right)^2\right\rbrack~.
\label{eq:asymp}
\eeq
We see that, as in the case of (\ref{eq:coth}),
 the late time asymptote for $V(\vphi)$ is once again the cosmological constant $V_0$. 
However the
presence of $\vphi^2$ in (\ref{eq:asymp}) suggests that the late time approach of $w_\vphi$ to $-1$ will be
oscillatory. This has been illustrated in figures \ref{fig:DE_coshw}, \ref{fig:DE_coshwwz} and \ref{fig:DE_coshww} which show $w_{\vphi}(z)$ and $\lbrace w_{\vphi},w_{\vphi}'\rbrace$
for different values of $\lambda$. 
From figure \ref{fig:DE_coshw},  it is clear that $w_{\vphi}=1/3,~0$ 
during the radiation and matter domination epochs respectively. However, at late times the scalar field 
begins to oscillate around the minimum of its potential (\ref{eq:asymp}). 
Since these oscillations are of decreasing amplitude, $w_{\vphi}$
asymptotically approaches $w_{\vphi} = -1$ at late times. 
Interestingly, for moderate values $5 \leq \lambda\leq 10$,  
the present value of $w_{\vphi}$ can lie anywhere between $-1$ to $-0.9$,
its precise value being determined by the phase of the oscillation. However 
for larger values $\lambda > 10$, the scalar field completes several oscillations prior to 
the present epoch.
Since the mean value of $\vphi(t)$ in (\ref{eq:asymp}) falls off as 
$\langle \vphi^2(t) \rangle^{1/2} \propto a^{-3/2}(t)$
it follows that in such models
 $w_{\vphi}\simeq -1$ today. 
This has been illustrated in fig. \ref{fig:DE_coshw} and especially in fig. \ref{fig:DE_coshwwz}. 

The following 
 expression describes the EOS of dark energy during the oscillatory epoch
\beq
w_{\vphi}(t)\simeq -1+\lambda^2 \left(\frac{\vphi_m(t)}{m_p}\right)^2 \left[1-\left(\frac{\vphi(t)}{\vphi_m(t)}\right)^2\right],
\label{eq:EOS2}
\eeq
where $\vphi_m(t)$ is the peak oscillation amplitude whose value steadily 
decreases with time.
Eq. (\ref{eq:EOS2}) can be rewritten as
\beq
w_{\vphi}(t)\simeq -1+\frac{\dot{\vphi}^2(t)}{V_0}~,
\label{eq:EOS3}
\eeq
where the steady decline of $\dot{\vphi}^2(t)$ with time ensures that
$w_{\vphi} \to -1$ at late times.

The previous analysis is substantiated by 
figure \ref{fig:DE_coshww} which shows the evolution of the phase-space 
$\lbrace w_{\vphi},w_{\vphi}'\rbrace$ for $\lambda=20,~30$ with filled black circles marking the 
present epoch ($\Omega_{0m}=0.3$). 
The substantial difference in $w_{\vphi}(z)$ for $5\leq \lambda \leq 30$ and 
$0 \leq z \leq 3$ (see fig. \ref{fig:DE_coshwwz}),
may allow such models to be differentiated from one another on the basis of the
high quality data expected from dark energy surveys such as DES, Euclid and SKA.

\begin{figure}[hbt]
\centering
\includegraphics[width=0.85\textwidth]{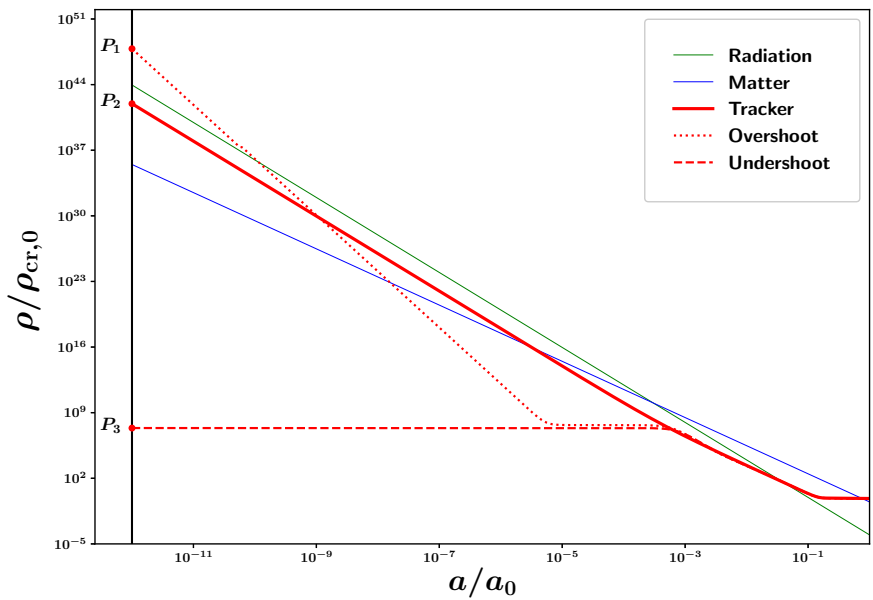}
\caption{The evolution of the scalar-field density
(red; in units of $\rho_{\rm cr,0}\equiv 3m_p^2H_0^2$) is shown together with the density in
 radiation (green) and matter (blue). 
The scalar field commences its descent from the exponential tracker asymptote of the potential $V(\vphi) = V_0\, \cosh{\left(\frac{\lambda\vphi}{m_p}\right)} $ with $\lambda=15$.  One finds that a very large range in initial scalar-field density values,
 covering over 40 orders of magnitude at $z \simeq 10^{12}$ ($P_{1}$ to $P_{3}$)
  leads to 
$\Omega_{0{\rm DE}} \simeq 0.7$ today. $P_{2}$
marks the initial density corresponding to the attractor solution (solid red)
to which all trajectories within the $P_{1}$--$P_{3}$ range converge by
$z \sim 100$.} \label{fig:DE_coshattractor}
\end{figure}

Due to the presence of the exponential tracker asymptote (\ref{eq:exp0}),  
the oscillatory tracker model
(\ref{eq:OLT1}) has a very large initial basin of attraction, trajectories from which get
funneled into the late time attractor $w_{\vphi} \simeq -1$. 
Our results, summarized in figure~\ref{fig:DE_coshattractor}, demonstrate that initial
density values covering a range of more than $40$ orders of magnitude at $z =
10^{12}$, converge onto the attractor scaling solution represented by
the solid red curve.
This range substantially increases if we set our initial
conditions at earlier times. For instance, upon setting $\{\vphi_i,\dot{\vphi}_i\}$ at the GUT
scale of $10^{14}\,{\rm GeV}$ ($z\sim 10^{26}$), the range of initial density values
that converges to $\Omega_{0,{\rm DE}}\simeq 0.7$ 
spans an impressive $100$ orders of magnitude\,!
The oscillatory tracker potential therefore exhibits a 
very large degree of freedom
in the choice of initial conditions. In particular it
permits the possibility of equipartition, according to which the density of dark energy
and radiation may have been comparable at very early times just after reheating. In our view this is a very compelling property of this DE model.\footnote{ See \cite{Roy:2013wqa,Paliathanasis:2015gga} for a  dynamical analysis of models based on similar potentials.}

In passing it may be appropriate to point out that the CPL ansatz \cite{polar,linder}
\beq
w(a) = w_0+w_1(1-a) = w_0 + w_1\frac{z}{1+z}~,
\label{eq:cpl}
\eeq
which is frequently used to reconstruct the properties of DE from observations,
may be unable to accommodate the oscillatory behaviour of $w_\vphi$ 
which characterizes this model. Non-parametric reconstruction is likely to work better
for this class of potentials \cite{DE1,non-param,woscillation1,woscillation2}.

\n
Note that the companion potential to (\ref{eq:OLT1}),
\beq
V(\vphi) = V_0 \sinh^{2}\frac{\tilde\lambda\vphi}{m_p}
\label{eq:DM}
\eeq
describes a tracker model of dark matter for 
$V_0\tilde\lambda^2/m_p^2 \gg H_0^2$ \cite{sss17}.
It is therefore interesting
 that, when taken together, the pair of $\alpha$-attractor potentials
(\ref{eq:OLT1}) and (\ref{eq:DM})
with $\tilde\lambda \gg \lambda$, can describe tracker models
of both dark matter {\em and} dark energy !

\subsection{The Recliner model}
\label{Exp_trac}

\begin{figure}[htb]
\centering
\includegraphics[width=0.8\textwidth]{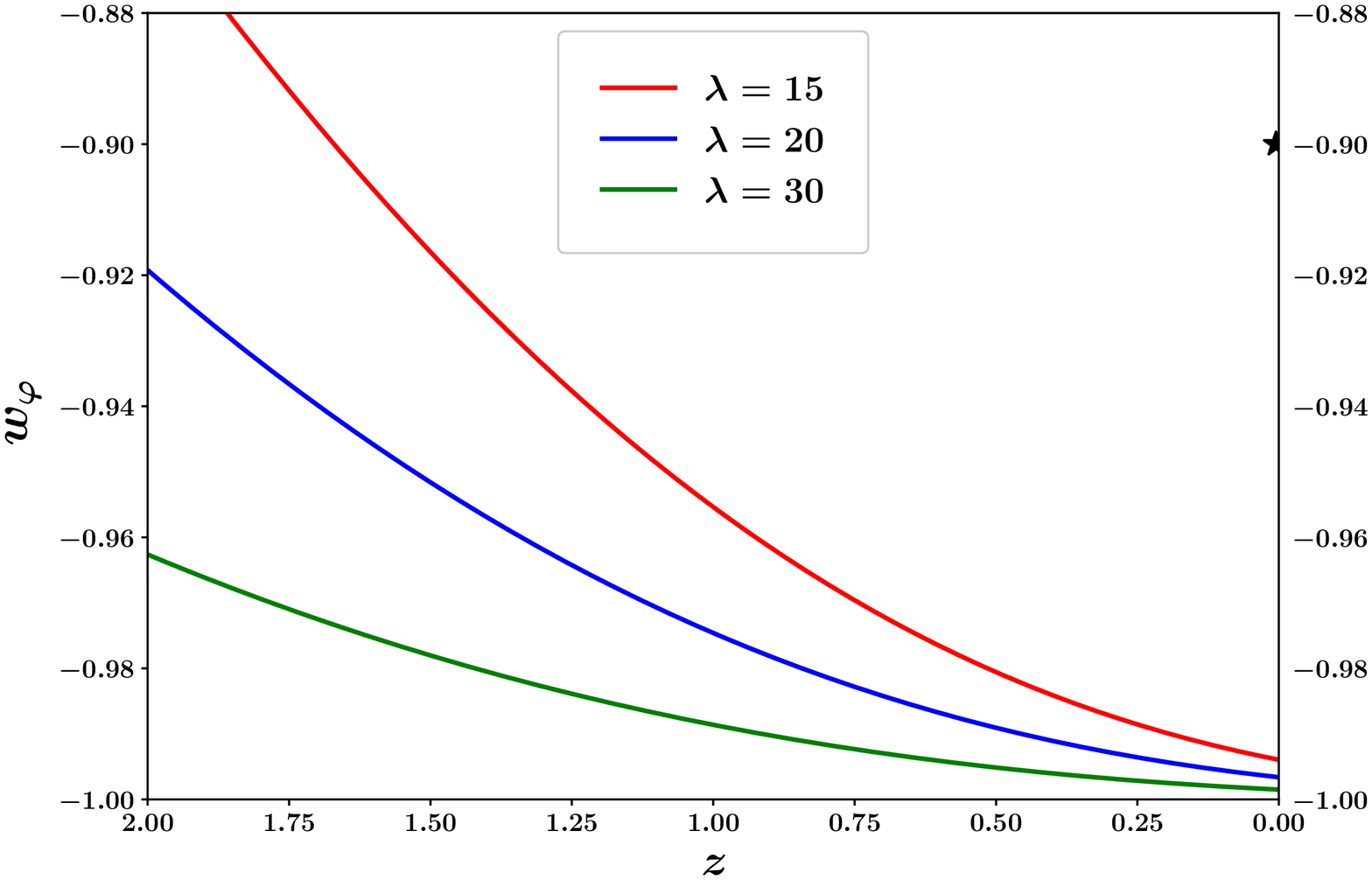}
\caption{The evolution of $w_{\vphi}(z)$ for the potential (\ref{eq:exp1})
is shown 
for $\lambda=15,~20,~30$. One notes that larger values of $\lambda$ 
lead to a more negative equation of state.
 The black star on the right  indicates the observational (conservative) upper bound on $w_0$ at $2 \sigma$ for DE models with a slowly varying EOS \cite{huterer17}.
} \label{fig:DE_star2RS}
\end{figure}

\begin{figure}[hbt]
\centering
\includegraphics[width=0.85\textwidth]{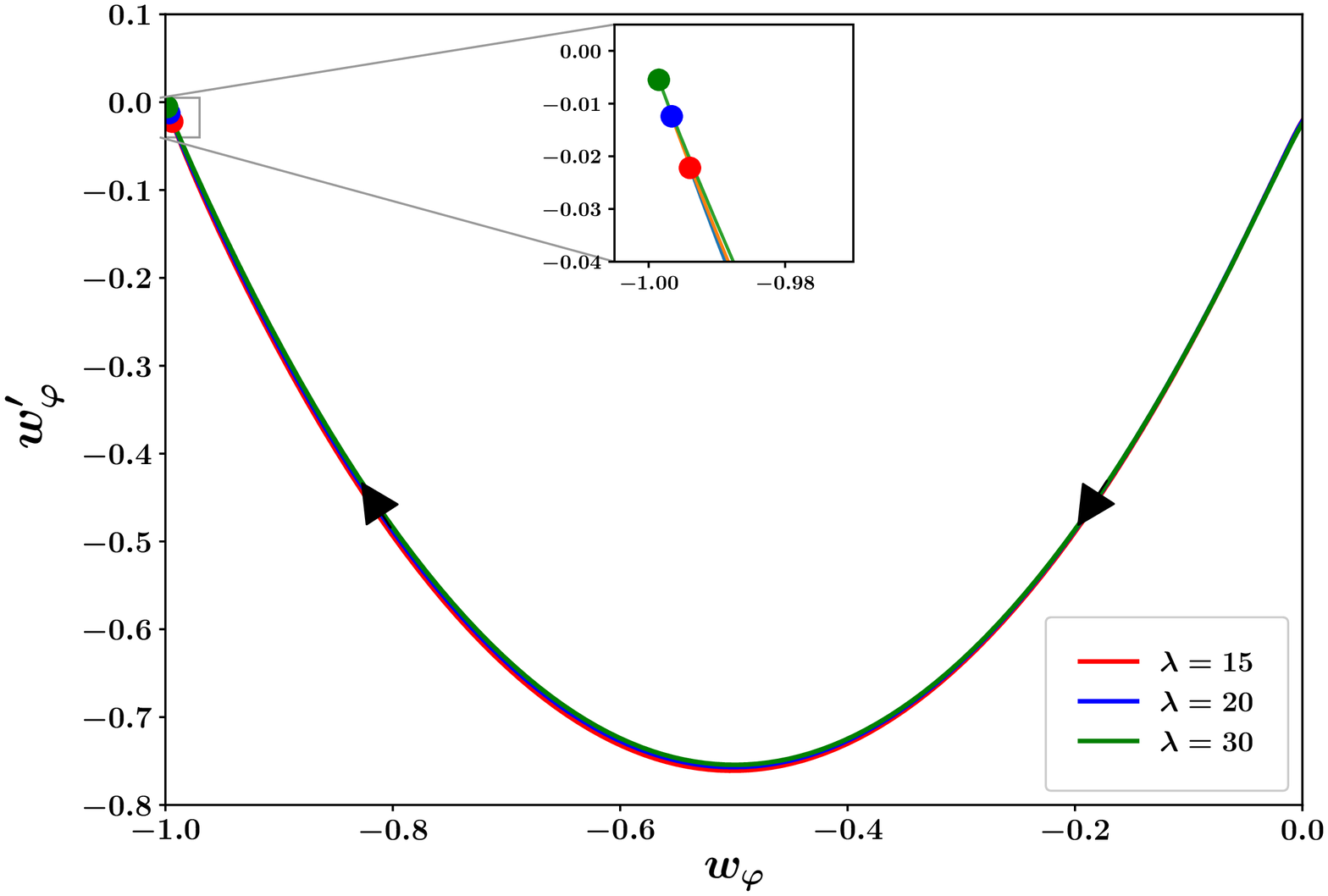}
\caption{Phase-space $\lbrace w_{\vphi},w_{\vphi}'\rbrace$ trajectories for the potential
$V(\vphi) =  V_0\, \left \lbrack 1+ \exp{(-\frac{\lambda\vphi}{m_p})}\right \rbrack$ 
are shown for different values of $\lambda$ starting from the matter dominated epoch. 
The large colored circles correspond to the present epoch ($z=0$) in each of the three 
models.
Arrows depict the flow of time.  
One finds that larger values of $\lambda$ result in smaller values of 
$w_{\vphi}$ and larger values of $w_{\vphi}'$ at late times.
} \label{fig:DE_expww}
\end{figure}

The recliner potential\footnote{The Recliner potential (\ref{eq:exp1}) presents a limiting
case of the family of potentials studied in \cite{barreiro}, also see \cite{scherrer_exp}.}
\beq
V(\vphi) = V_0\, \left \lbrack 1+ \exp{(-\frac{\lambda\vphi}{m_p})}\right \rbrack~,
\label{eq:exp1}
\eeq
possesses the asymptotic form $V \simeq V_0\exp{(-\frac{\lambda\vphi}{m_p})}$
for $\lambda|\vphi| \gg m_p$ ($\vphi < 0$).
This endows it with a  large initial basin of attraction, due to which 
scalar field trajectories rolling down (\ref{eq:exp1}) 
approach a common evolutionary path from a wide range of initial conditions.
\begin{figure}[hbt]
\centering
\includegraphics[width=0.83\textwidth]{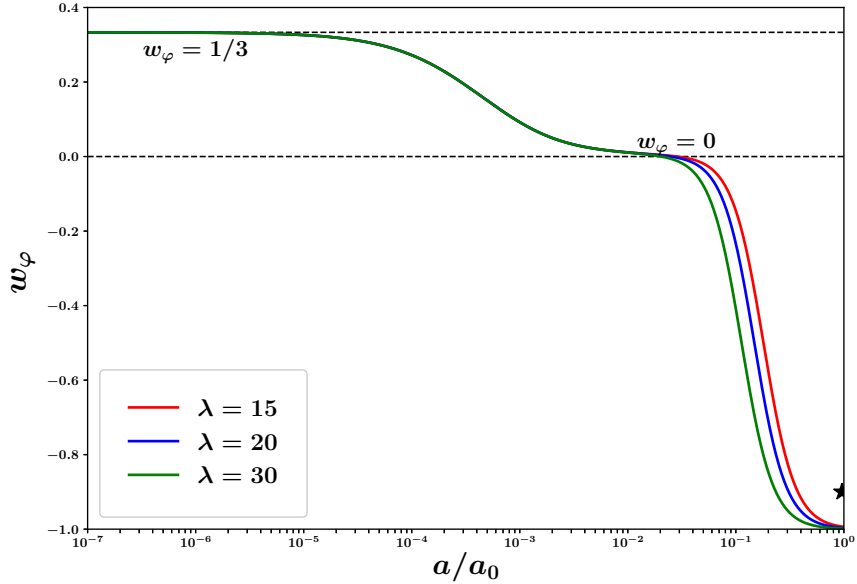}
\caption{This figure illustrates the evolution of $w_{\vphi}$ for the
 potential $V(\vphi) = V_0\, \left \lbrack 1+ \exp{(-\frac{\lambda\vphi}{m_p})}\right \rbrack$, 
for different values of $\lambda$. One finds that during the radiation and matter dominated 
epochs, when $V \simeq V_0 \exp{\left(-\frac{\lambda\vphi}{m_p}\right)}$, the scalar field tracks the cosmological background density with $w_{\vphi}=1/3,~0$ respectively. 
However at  late times  the value of $w_{\vphi}$ sharply drops and approaches 
$w_{\vphi} = -1$ asymptotically. Larger values of $\lambda$ result
in more negative values of $w_{\vphi}$ at late times.
 The black star on the right  indicates the (conservative) observational 
upper bound on $w_0$ (at $2 \sigma$) for DE models with a slowly varying EOS \cite{huterer17}.}
\label{fig:DE_expw}
\end{figure}
In the  large $\vphi$ limit, $\lambda\vphi \gg m_p$, $V(\vphi)$ monotonically 
 declines to $V(\vphi)\simeq V_0$. Therefore one expects the value of
$w_{\vphi}$ to approach $w_{\vphi}\to -1$ at late times, without any intermediate
oscillations. This behaviour is substantiated by figure \ref{fig:DE_expw}. 
The fact that the current value of $w_{\vphi}$ can fall below $-0.9$ 
makes this model quite appealing since it can describe cosmic acceleration 
without the fine tuning
of initial conditions. The evolution of $w_{\vphi}(z)$ near the present epoch is
shown in figure \ref{fig:DE_star2RS}.
One finds that different values of $\lambda$ in (\ref{eq:exp1}) can clearly be distinguished
on the basis of low redshift measurements of $w_{\vphi}(z)$.
Consequently upcoming dark energy survey's (DES, Euclid, SKA, etc.) may provide
a unique opportunity to set bounds (or even determine) the value of $\lambda$
in (\ref{eq:exp1}).
Phase-space trajectories $\lbrace w_{\vphi},w_{\vphi}'\rbrace$ for 
the Recliner potential (\ref{eq:exp1}) are 
illustrated in 
figure \ref{fig:DE_expww}. 
Note that larger values of $\lambda$ in (\ref{eq:exp1}) result in smaller values of
$w_{\vphi}$ and larger values of $w_{\vphi}'$ at late times.

It is interesting that the
Oscillating tracker potential (\ref{eq:OLT1}) and the Recliner potential(\ref{eq:exp1})
have two free parameters each, $V_0$ and $\lambda$, precisely the same number 
as in the IPL model $V = V_0\vphi^{-p}$ and the tracker $V = V_0 \exp{\left(-\frac{\lambda\vphi}{m_p}\right)}$.

\subsection{Transient dark energy from the Margarita potential}
\label{Trans_DE}
\begin{figure}[ht]
\centering
\includegraphics[width=0.65\textwidth]{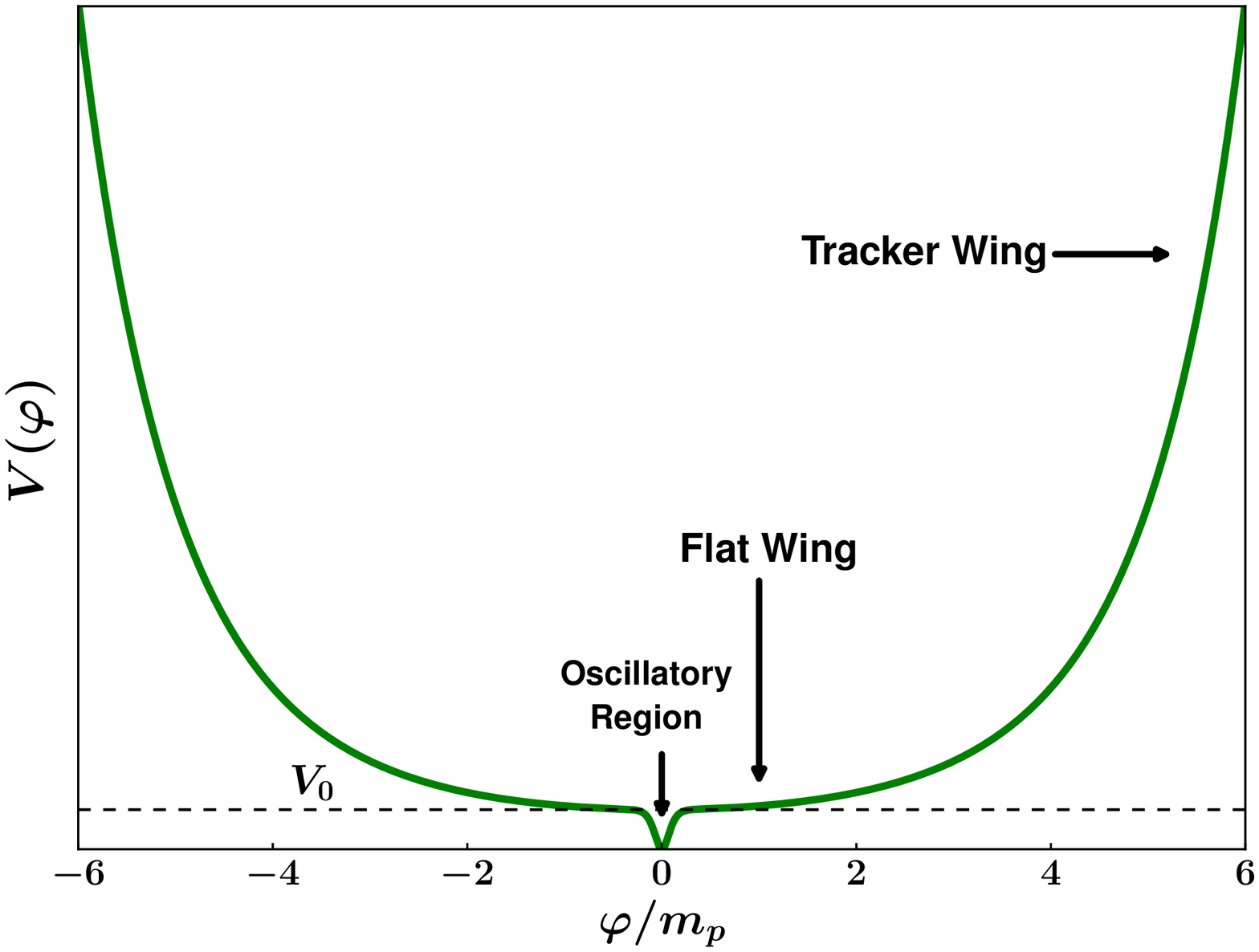}
\caption{This figure schematically illustrates the transient DE  potential (\ref{eq:champagne1})
 with $\lambda_1=10$ and $\lambda_2=1$.
The main features of this potential are: 
 exponential tracker wings for $\frac{|\vphi|}{m_p} \gg \frac{1}{\lambda_2}$,
flat wings for $\frac{1}{\lambda_1}\ll \frac{|\vphi|}{m_p} \ll \frac{1}{\lambda_2}$, 
and an oscillatory region ($\frac{|\vphi|}{m_p} \ll \frac{1}{\lambda_1}$)
where $V \propto \vphi^2$. }
\label{fig:champagne}
\end{figure}

 \begin{figure}[hbt]
\centering
\includegraphics[width=0.85\textwidth]{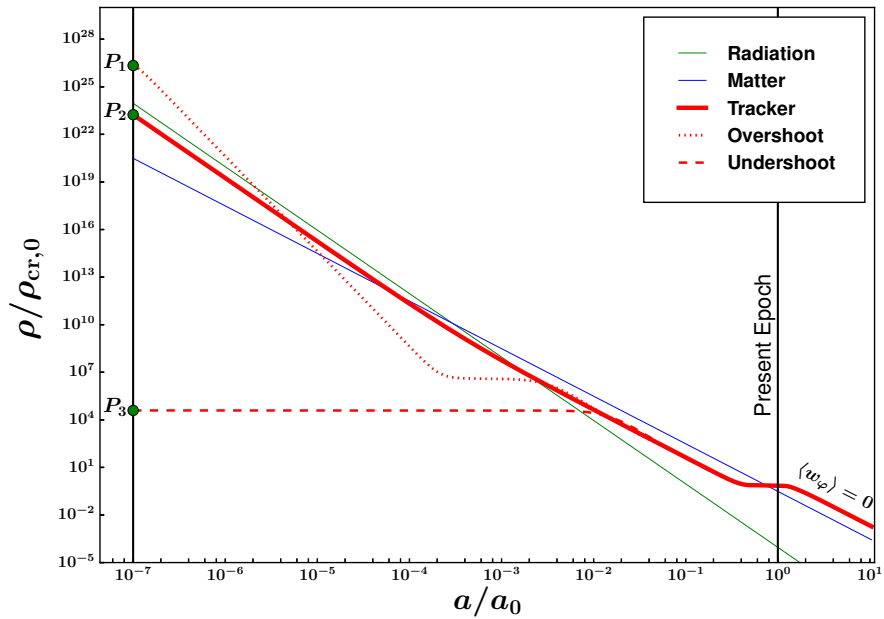}
\caption{The evolution of the scalar-field density (red), in units 
of $\rho_{\rm cr,0}=3m_p^2H_0^2$, is shown from $z
\simeq 10^{7}$ until $z = 0$. The scalar field commences its descent from the exponential tracker asymptote of the potential (\ref{eq:champagne1})
with $\lambda_1=500$ and $\lambda_2=5$.  
A large range in initial (scalar-field) density values, covering more than 20 orders of magnitude from $P_{1}$ to $P_{3}$,  leads to 
$\Omega_{0,{\rm DE}} \simeq 0.7$ today. $P_{2}$
marks the initial density corresponding to the attractor solution (solid red)
to which all trajectories commencing in the $P_{1}$--$P_{3}$ range converge. 
Note that cosmic acceleration 
 is a transient phenomenon since the universe reverts to matter domination in the
 future once the scalar field begins to oscillate with $\langle w_\vphi\rangle = 0$. } \label{fig:DE_champattractor}
\end{figure}

\begin{figure}[hbt]
\centering
\subfigure[][]{
\includegraphics[width=0.6\textwidth]
{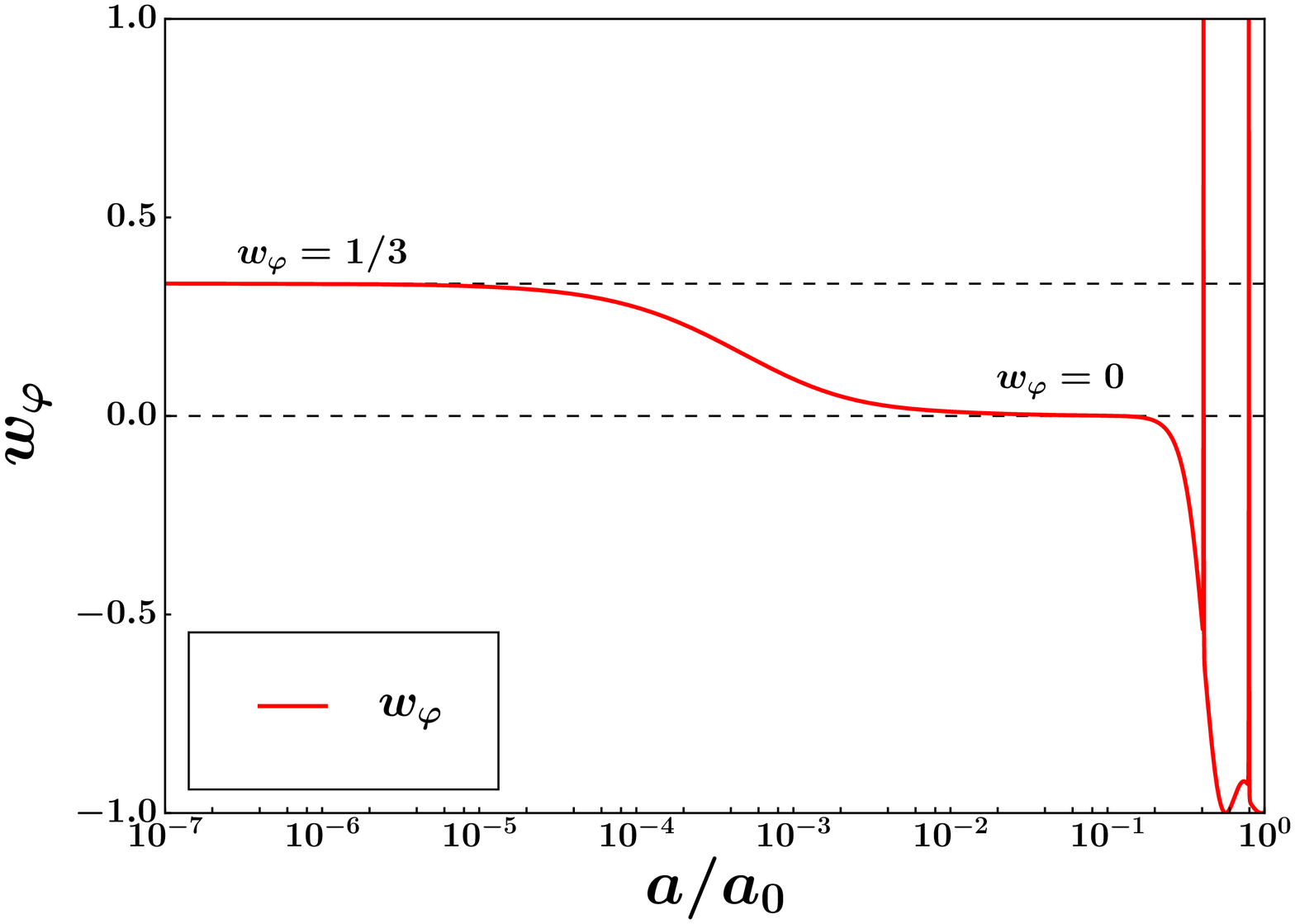}\label{fig:DE_champw}}
\subfigure[][]{
\includegraphics[width=0.6\textwidth]
{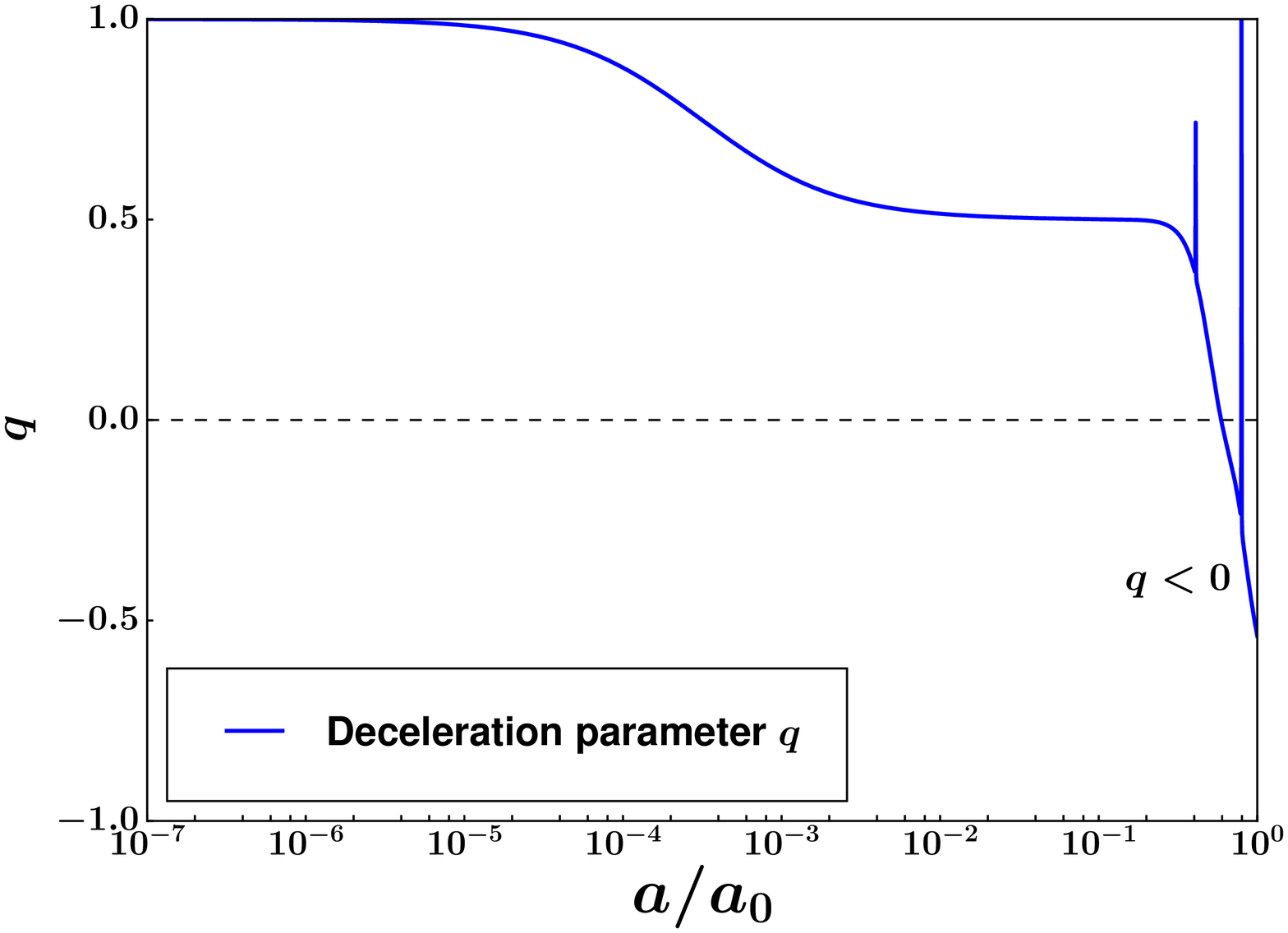}\label{fig:DE_champq}}
\caption{The evolution of $w_{\vphi}$ (upper panel) and the deceleration parameter 
$q=-\frac{\ddot a}{aH^2}$ (lower panel) is shown for  the transient DE model 
(\ref{eq:champagne1}) with $\lambda_1=500$ and $\lambda_2=5$. One finds that 
during the radiation and matter dominated epochs the scalar field tracks the 
cosmological background density with $w_{\vphi}=1/3,~0$ respectively. 
However at  late times the scalar field gets trapped within the flat wing of the
Margarita potential shown in fig. \ref{fig:champagne}. When this happens $\vphi(t)$
begins to oscillate within the flat wing and $w_{\vphi}$ drops to negative
values, signifying cosmic acceleration. It is interesting that the oscillatory epoch
is punctuated by sharp spikes in $w_{\vphi}$ and $q$. Spikes occur whenever $\vphi(t)$
ventures inside the sharp potential well at $\vphi(t) \simeq 0$, see fig. \ref{fig:champagne}.
When this happens $w_{\vphi}$ and $q$ abruptly increase and, for a very brief 
period of time,
acceleration $q<0$ gives way to deceleration $q>0$.
A magnified view of this region is shown in the next figure.
Finally, in the distant future, $\vphi$ gets trapped within the {\em oscillatory region}
near $\vphi \simeq 0$ shown in fig. \ref{fig:champagne}.
At this point rapid oscillations of $\vphi(t)$ ensure that the effective equation of state
is pressureless, $\langle w_{\vphi} \rangle = 0$, 
and the universe expands as if matter dominated; see fig. \ref{fig:DE_champw1}.
}
\label{fig:DE_champqw}
\end{figure}

The Margarita potential describes a model of transient dark energy
\beq
V(\vphi) = V_0 \tanh^2{\left(\frac{\lambda_1\vphi}{m_p}\right)}\cosh{\left(\frac{\lambda_2\vphi}{m_p}\right)}, ~~~~\lambda_1 \gg \lambda_2~.
\label{eq:champagne1}
\eeq

This  potential has tracker-like wings and
 a flat intermediate region (see figure~\ref{fig:champagne}).
It exhibits three asymptotic branches: 
\ber
\mbox{Tracker wing:} \quad   V(\vphi) &\simeq& \frac{V_0}{2}\, \exp{\left(\lambda_2|\vphi|/m_p\right)}~,  \quad \frac{|\vphi|}{m_p} \gg \frac{1}{\lambda_2}\, ,
\label{eq:champpot1}\\
\mbox{Flat wing:} \quad V(\vphi) &\simeq& V_0+\frac{1}{2}m_2^2\vphi^2\, ,\quad \frac{1}{\lambda_1}\ll \frac{|\vphi|}{m_p} \ll \frac{1}{\lambda_2}\, ,
\label{eq:champpot2}\\
\mbox{Oscillatory region:} \quad V(\vphi) &\simeq& \frac{1}{2}m_1^2\vphi^2\, , \quad \frac{|\vphi|}{m_p} \ll \frac{1}{\lambda_1}\, ,
 \label{eq:champpot3}
\eer
 where $m_1^2=\frac{2V_0 \lambda_1^2}{m_p^2}$ and $m_2^2=\frac{V_0 \lambda_2^2}{m_p^2}$ with $m_1\gg m_2$.

\begin{figure}[hbt]
\centering
\includegraphics[width=0.75\textwidth]{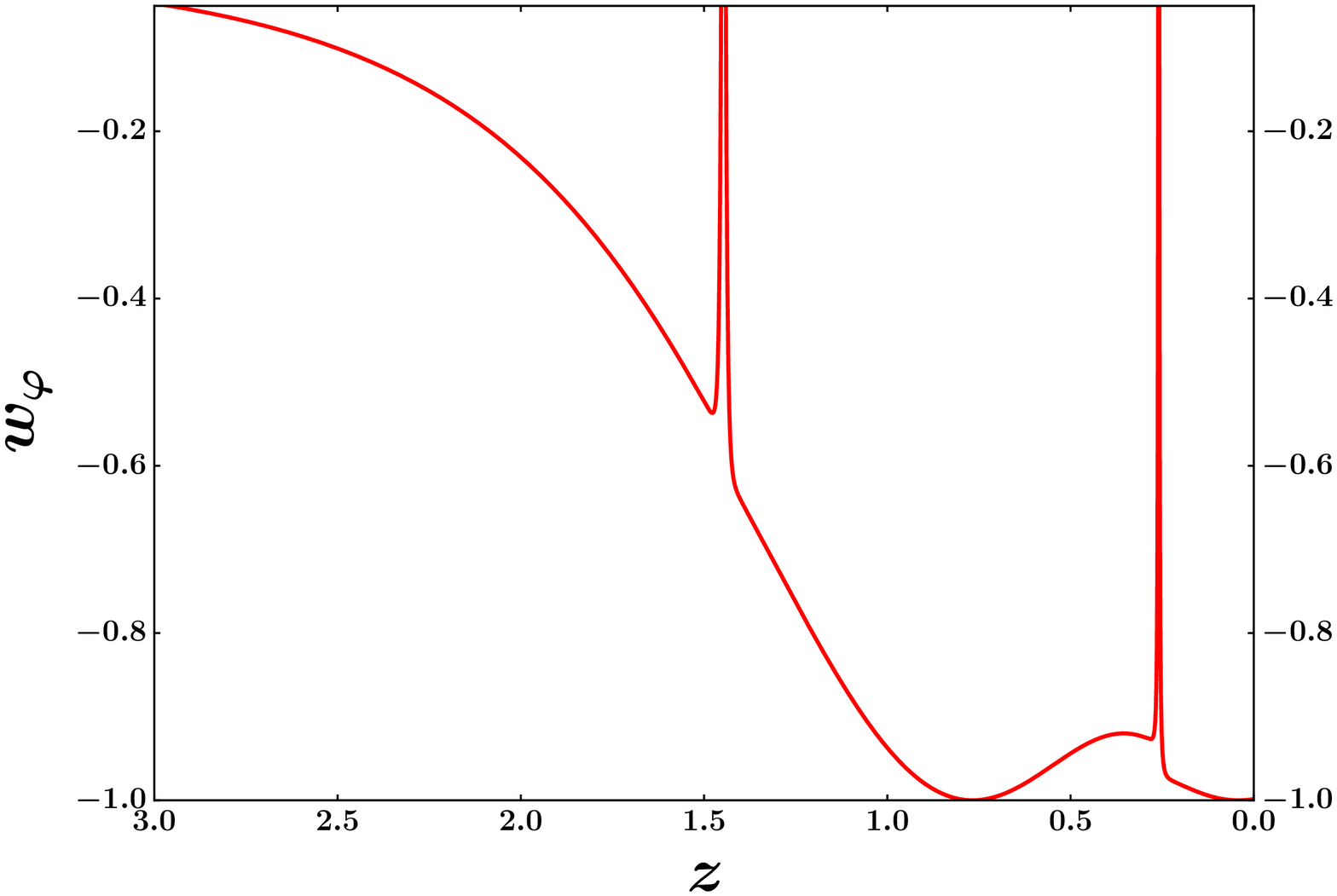}
\caption{The presence of spikes in the EOS of dark energy is highlighted by this figure.
Spikes arise whenever $\vphi(t)$ traverses the region near the origin
$\vphi \simeq 0$  where $V(\vphi)$ has a
sharp feature; see fig. \ref{fig:champagne}.
}
\label{fig:DE_zoom}
\end{figure} 

As illustrated by the red curve in figure \ref{fig:DE_champattractor}, the acceleration of
the universe in this model is a transient phenomenon. It ends once the
scalar field rolls to the minimum of $V(\vphi)$.
From that point on the scalar field begins to oscillate
and behave like dark matter.
Consequently the universe reverts to matter
dominated expansion after the current accelerating epoch is over,
with an extra contribution to dark matter coming from the coherently oscillating scalar field.

\begin{figure}[ht]
\centering
\includegraphics[width=0.75\textwidth]{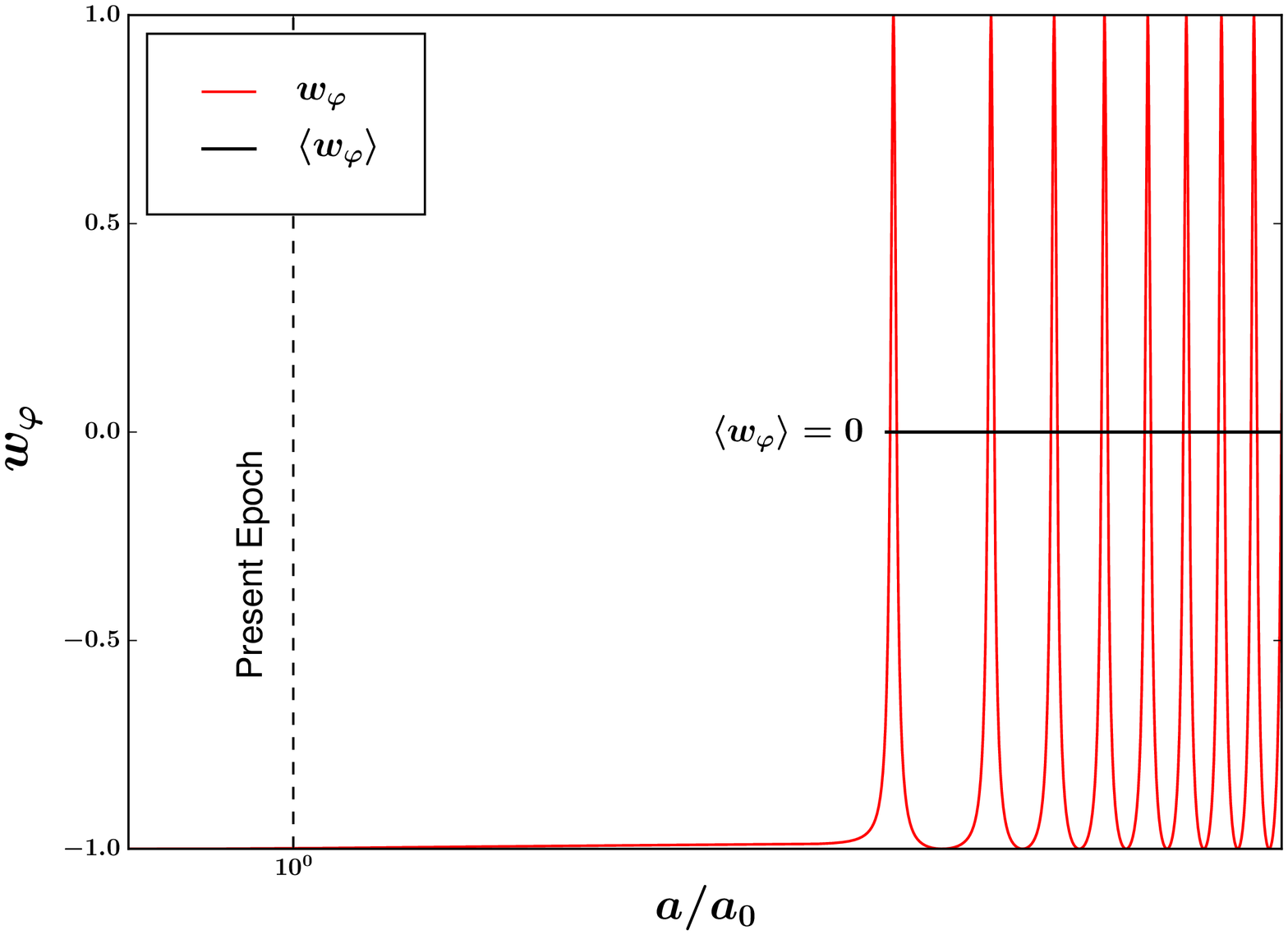}
\caption{The future evolution of  $w_{\vphi}$ is shown for the transient DE model 
(\ref{eq:champagne1}). Note that the scalar field behaves almost
like a cosmological constant 
($w_{\vphi}\simeq -1$) close to the present epoch.
 At late times rapid oscillations of $\vphi(t)$ (within the {\em oscillatory region} of
 figure \ref{fig:champagne}) result in
the scalar field behaving like pressureless (dark) matter with
 $\left\langle w_{\varphi}\right\rangle = 0$.
Thus the present accelerating epoch is sandwiched between two matter dominated epochs.
}
\label{fig:DE_champw1}
\end{figure}

As suggested by
 (\ref{eq:champpot1}) -- (\ref{eq:champpot3}), the motion of the scalar field
proceeds along three distinct stages each of which is
 reflected in the cosmic expansion history.

(i) Initially $\vphi(t)$ rolls down the exponential potential (\ref{eq:champpot1}).
During this phase the scalar field density scales like the background fluid (radiation/matter)
driving the expansion of the universe.
(ii) 
After the  tracking phase is over, the  scalar field oscillates around the 
{\em flat wing} of the potential shown in figure \ref{fig:champagne}.
During this phase the universe begins to accelerate, as demonstrated in
figure \ref{fig:DE_champqw}. (iii) Finally, at late times, 
the scalar field gets trapped within the sharp {\em oscillatory region} of the potential 
(\ref{eq:champpot3}); see figure \ref{fig:champagne}.
Oscillations of the scalar field during this stage 
 make it behave like pressureless matter with  $\left\langle w_{\varphi}\right\rangle=0$;
 see the red line in figures \ref{fig:DE_champattractor} and \ref{fig:DE_champw1}.   

We therefore find that cosmic acceleration is sandwiched between
two matter dominated epochs. The duration of the accelerating phase depends upon
 the gap between $\lambda_1$ and $\lambda_2$. Nucleosynthesis constraints limit $\lambda_2\geq 5$ 
whereas the only constraint on $\lambda_1$ comes from the inequality $\lambda_1 \gg \lambda_2$.

At this point we would like to draw attention to a key feature of the 
Margarita potential which distinguishes this model of transient acceleration
from others of its kind.
Note that the asymptotic form of the
 potential within the flat wing, described by (\ref{eq:champpot2}),
bears close resemblance to the potential near $\vphi \simeq 0$ for the oscillatory
tracker model, namely (\ref{eq:asymp}). Therefore, as in that model,
 one might expect $w_\vphi$ to approach $-1$ 
at late times via small oscillations.
This would indeed be the case were it not for the presence of the sharp 
{\em oscillatory region}
near $\vphi \simeq 0$ in fig. \ref{fig:champagne}.
This region modifies the behaviour of $\vphi$ significantly.
As $\vphi$ traverses $\vphi \simeq 0$, its EOS abruptly changes from negative to
positive values. This leads to a spike in the value of $w_\vphi$ and in the deceleration
parameter $q$. An accelerating universe punctuated by periods of sudden deceleration
therefore becomes a key feature of the Margarita model of DE, as shown in
figures \ref{fig:DE_champqw} \& \ref{fig:DE_zoom}. 

It is well known that, unlike $\Lambda$CDM, a transiently accelerating universe  does not 
possess a 
future event horizon. Moreover the presence of even a tiny curvature term,
$k/a^2$, can cause such a universe to stop expanding and begin to contract,
giving rise to a very different cosmological future from $\Lambda$CDM.
Note that while models of transient DE have been discussed earlier, 
see for instance \cite{transient,ss02,star17}, to the best of our
knowledge none of these early models had a tracker-like
 commencement.

Finally the reader may be interesting in a 
companion potential to (\ref{eq:champagne1}) which provides
another example of transient dark energy with tracker-like 
 behaviour at early times
\beq
V(\vphi) = V_0 \left(1-e^{-\left(\frac{\lambda_1\vphi}{m_p}\right)^2}\right)\cosh{\left(\frac{\lambda_2\vphi}{m_p}\right)}, ~~~~\lambda_1\gg\lambda_2~.
\label{eq:champagne2}
\eeq

\section{Discussion}
\label{sec:Summary}
In this paper we discuss four new models of dark energy based on the
$\alpha$-attractor family.
In all of these models 
the present value of the equation of state
 can fall below $-0.9$, in agreement with recent
observations.
This does not come at the expense of finely tuned initial conditions, since all of
the four models display tracker-like behaviour at early times.
The initial attractor basin is largest for 
the Oscillatory tracker model (\ref{eq:OLT}),
the Recliner potential (\ref{eq:exp})
and
the Margarita potential (\ref{eq:champagne}), in all of which 
$V \propto e^{\lambda\vphi}$
at early times. The fourth model, which is described by
the L-potential (\ref{eq:PLT}),
has exactly the same basin of attraction as the inverse power law
potential $V \propto \vphi^{-p}$. 
It is interesting that all of these models display distinct
late time features which allow them to be easily distinguished from one another.
For instance, in the Oscillatory tracker model (\ref{eq:OLT}) the late-time
attractor $w_\vphi \simeq -1$ 
is reached through a series of oscillations of decreasing amplitude.
By contrast, oscillations in
$w_\vphi$ are absent in the Recliner potential (\ref{eq:exp}), 
in which the late-time approach to
$w_\vphi \simeq -1$ occurs via a steady decline in the value of $w_\vphi$. 

Our fourth model, represented by the
Margarita potential (\ref{eq:champagne}), describes a {\em transient} model
of dark energy.
In this model the accelerating phase is sandwiched between two matter 
dominated epochs.
However unlike other transiently accelerating
 models discussed in the literature, the Margarita potential
provides us with an example of an $\alpha$-attractor based model with a
tracker-like asymptote at early times. This ensures that transient 
acceleration
can arise from a fairly large family of initial conditions. 

Finally we would like to mention that the potentials suggested in this paper do not claim to address
the `why now' question which is sometimes raised in the context of dark energy.
The potentials in our paper contain two free parameters $V_0$ and $\lambda$.
The value of $\lambda$ is chosen in keeping with the requirement that the EOS can drop to
the low
values demanded by observations \cite{huterer17}.  The value of the other free parameter $V_0$ is adjusted to
ensure $\Omega_m \simeq 1/3$, $\Omega_{\rm DE} \simeq 2/3$ at the present epoch.
Its important to note that 
for a given value of $V_0$ there is an entire range of
initial conditions $\lbrace \vphi_i, {\dot \vphi}_i\rbrace$ which funnel dark energy to its present value.
This ensures that there is little fine tuning of initial conditions in the models discussed in this paper.

\section*{Acknowledgments}
The authors acknowledge useful discussions with Yu. Shtanov and A. Viznyuk. S.B. and S.S.M. thank the Council of Scientific and Industrial Research (CSIR), India, for financial support as senior research fellows.

\appendix

\section{Can \lcdm~ cosmology emerge from a single oscillatory potential ?}
\label{sec:unity}

Consider any potential such as $V = V_0 \cosh{\lambda\vphi/m_p}$ in 
\eqref{eq:OLT1} which has an early time tracker phase 
and  the late time asymptotic form 
\beq
V(\varphi) \simeq V_0\left[1+\frac{1}{2}\left(\frac{\lambda\vphi}{m_p}\right)^2 \right],
~~~~ |\lambda\vphi|\ll m_p~.
\label{eq:A1}
\eeq
Recasting \eqref{eq:A1} as
\beq
V(\varphi) \simeq V_0+\frac{1}{2}m^2\vphi^2,
\label{eq:A2}
\eeq
where $m^2=\frac{V_0\lambda^2}{m_p^2}$, 
one might be led into thinking that:
(i) since $V_0$ behaves like the
cosmological constant, and (ii) 
 the $m^2\vphi^2$ term leads to oscillations in $\vphi$ during which
$\langle w_\vphi\rangle \simeq 0$, therefore
a potential having the general asymptotic form \eqref{eq:A2} might be able to play the dual role
of describing both
 dark matter and dark energy. However this is not the case for the simple
reason that although
oscillations commence when $m\ggeq H$, 
which  might lead one to believe that $\frac{1}{2}m^2\vphi^2 \gg V_0$,
the asymptotic forms (\ref{eq:A1}) and 
(\ref{eq:A2}) are only valid in the limit $|\lambda\vphi|\ll m_p$ 
which implies $\frac{1}{2}m^2\vphi^2 \ll V_0$. In other words,
the cosmological constant $V_0$ is always larger than the oscillatory $m^2\vphi^2$ term
soon after the onset of oscillations, leaving little room for
a prolonged dark matter dominated epoch as demanded by observations.

Note, however, that a viable model of \lcdm~ based on a single scalar field can be
constructed within the framework of a non-canonical Lagrangian, as shown in
\cite{sahni-sen17}.


\begin{thebibliography}{99}

\bibitem{ss00}
V. Sahni and A.A. Starobinsky, Int. J. Mod. Phys. {\bf D9} 373 (2000).

\bibitem{DE1}
V. Sahni and A.A. Starobinsky, Int. J. Mod. Phys. {\bf D15} 2105 (2006).

\bibitem{DE}
P.~J.~E. Peebles and B. Ratra, Rev. Mod. Phys. {\bf 75} 559 (2003);
T. Padmanabhan, Phys. Rep. {\bf 380} 235 (2003);
V. Sahni, [astro-ph/0202076], [astro-ph/0502032];
V. Sahni, {\em Dark matter and dark energy}, Lect. Notes Phys. 653, 141-180 (2004)  [astro-ph/0403324].
E. J. Copeland, M. Sami and S. Tsujikawa, Int. J. Mod. Phys. {\bf D15} 1753 (2006);
R. Bousso, Gen. Relativ. Gravit. {\bf 40}, 607 (2008);
L. Amendola and S. Tsujikawa, {\em Dark Energy}, Cambridge University Press, 2010.

%

\bibitem{ratra}
B. Ratra and P.J.E. Peebles, \prd {\bf 37}, 3406 (1988).

\bibitem{wetterich88}
C. Wetterich, Nuclear Physics B {\bf 302}, 668 (1988).

\bibitem{ferreira}
P.G. Ferreira and M. Joyce, \prl {\bf 79}, 4740 (1997);
P.G. Ferreira, P.G. and M. Joyce, \prd {\bf 58}, 023503 (1998).

\bibitem{zlatev}
I. Zlatev L. Wang and P.J. Steinhardt, \prl {\bf 82}, 896 (1999).

\bibitem{zlatev1}
P.J. Steinhardt, L. Wang and I. Zlatev, \prd {\bf 59}, 123504 (1999).

\bibitem{sw00}
V. Sahni and L. Wang, \prd {\bf 62}, 103517 (2000).

\bibitem{brax}
P. Brax and J. Martin, \prd {\bf 61}, 103502 (2000);
\plb {\bf 468}, 40 (1999).

\bibitem{barreiro}
T. Barreiro, E.J. Copeland and N.J. Nunes, \prd {\bf 61}, 127301 (2000).

\bibitem{albrecht00}
A. Albrecht and C. Skordis, \prl {\bf 84}, 2076 (2000).

\bibitem{BAO}
Y. Wang et al., SDSS Collaboration, MNRAS {\bf 469}, 3762 (2017) [arXiv:1607.03154].

\bibitem{SDSS}
S. Alam et al., BOSS Collaboration, MNRAS {\bf 470}, 2617 (2017)
[arXiv:1607.03155].

\bibitem{planck}
P.A.R. Ade et al., Planck Collaboration, \asta{\bf 594}, A14  (2016),  Dark energy and modified gravity, [arXiv:1502.01590].

\bibitem{asen17}
A. I. Lonappan, Ruchika, A. A Sen, arXiv:1705.07336.

\bibitem{huterer17}
D. Huterer and D.L. Shafer, Rep. Prog. Phys. {\bf 81}, 016901 (2018) [arXiv:1709.01091].

\bibitem{linde1}
R. Kallosh and A. Linde, JCAP07 (2013) 002 [arXiv:1306.5220].

\bibitem{linde2}
R. Kallosh, A. Linde and D. Roest, JHEP11, 198 (2013) [arXiv:1311.0472].

\bibitem{sss17}
S. Mishra, V. Sahni and Yu. Shtanov, JCAP 06(2017)045 [arXiv:1703.03295]


\bibitem{linder_alpha}
E. V. Linder, \prd {\bf 91}, no. 12, 123012 (2015) [arXiv:1505.00815].

\bibitem{chap}
A. Kamenshchik, U. Moschella and V. Pasquier, 2001, \plb {\bf 511}, 265
[{\tt gr-qc/0103004}].

\bibitem{chap1}
V. Gorini, A. Kamenshchik, U. Moschella, V. Pasquier and A. Starobinsky,
2005, \prd {\bf 72}, 103518 [{\tt astro-ph/0504576}].

\bibitem{bilic}
N. Bilic, G.B. Tupper and R. Viollier, 2002, \plb {\bf 535}, 17
[{\tt astro-ph/0111325}].

\bibitem{frolov}
A. Frolov, L. Kofman and A. Starobinsky, 2002, \plb {\bf 545}, 8
[{\tt hep-th/0204187}].


\bibitem{matarrese}
S. Matarrese, C. Baccigalupi and F. Perrotta, \prd {\bf 70}, 061301 (2004).

\bibitem{caldwell_linder}
R.R. Caldwell and E.V. Linder, \prl {\bf 95}, 141301 (2005).

\bibitem{linder17}
E.V. Linder, Astropart. Phys. {\bf 91}, 11 (2017) [arXiv:1701.01445].

\bibitem{sfs}
V. Sahni, H. Feldman and A. Stebbins, Astrophys.J. {\bf 385}, 1, (1992).

\bibitem{copeland98}
E.J. Copeland, A.R. Liddle and D. Wands, \prd {\bf 57}, 4686 (1998).

\bibitem{CMB_Neff1}
E.~Calabrese, D.~Huterer, E.~V.~Linder, A.~Melchiorri and L.~Pagano,
  \prd {\bf 83}, 123504 (2011) [arXiv:1103.4132 [astro-ph.CO]].

\bibitem{CMB_Neff2}
 A.~Hojjati, E.~V.~Linder and J.~Samsing,
  \prl  {\bf 111},  041301  (2013) [arXiv:1304.3724 [astro-ph.CO]].

\bibitem{scherrer_exp}
H. Chang, R. J. Scherrer, [arXiv:1608.03291].

\bibitem{Roy:2013wqa} 
  N.~Roy and N.~Banerjee,
  Gen.\ Rel.\ Grav.\  {\bf 46}, 1651 (2014)
  [arXiv:1312.2670 [gr-qc]].
  
\bibitem{Paliathanasis:2015gga} 
  A.~Paliathanasis, M.~Tsamparlis, S.~Basilakos and J.~D.~Barrow,
  Phys.\ Rev.\ D {\bf 91}, no. 12, 123535 (2015)
  [arXiv:1503.05750 [gr-qc]].  


\bibitem{polar}
M. Chevallier and D. Polarski, Int. J. Mod. Phys. D {\bf 10}, 213 (2001)
[{\tt gr-qc/0009008}].

\bibitem{linder}
E. V. Linder, \prl {\bf 90}, 091301 (2003)
[{\tt astro-ph/0208512}].

\bibitem{non-param}
A. Shafieloo, U. Alam, V. Sahni, and A.A. Starobinsky, Mon.
Not. R. Astron. Soc. {\bf 366}, 1081 (2006); 
J. Dick,L. Knox, and M. Chu, J. Cosmol. Astropart. Phys. 07 (2006)
001; A. Shafieloo, Mon. Not. R. Astron. Soc. {\bf 380}, 1573
(2007);
 D. Huterer and G. Starkman, Phys. Rev. Lett. {\bf 90}, 031301
(2003);
R. G. Crittenden, L. Pogosian, and G. B. Zhao, J. Cosmol.
Astropart. Phys. 12 (2009) 025;
C. Clarkson and C. Zunckel, Phys. Rev. Lett. {\bf 104}, 211301
(2010);
R. G. Crittenden, G. B. Zhao, L. Pogosian, L. Samushia,
and X. Zhang, J. Cosmol. Astropart. Phys. 02 (2012)
048;
T. Holsclaw, U. Alam, B. Sanso´, H. Lee, K. Heitmann,
S. Habib, and D. Higdon, Phys. Rev. Lett. {\bf 105}, 241302
(2010);
A. Shafieloo, A. G. Kim, and E.V. Linder, Phys. Rev. D {\bf 85}, 123530 (2012);
M. Seikel, C. Clarkson, and M. Smith, J. Cosmol.
Astropart. Phys. 06 (2012) 036.

\bibitem{woscillation1}
G-B. Zhao, R. G. Crittenden, L. Pogosian and X. Zhang,
\prl {\bf 109}, 171301 (2012).

\bibitem{woscillation2}
G-B Zhao, et. al., Nature Astronomy {\bf 1}, 627 (2017).










\bibitem{transient}
J. Frieman, C.T. Hill, A. Stebbins and I. Waga, \prl {\bf 75},
2077 (1995);
K. Choi, Phys.Rev. D {\bf 62} 043509 (2000) [hep-ph/9902292];
J.D. Barrow, R. Bean, and J. Magueijo, MNRAS {\bf 316}, L41 (2000);
S.C. Ng and D.L. Wiltshire, 
Phys.Rev. D {\bf 64} 123519 (2001) [astro-ph/0107142];
R. Kallosh, A. Linde, S. Prokushkin and M. Shmakova, 
\prd {\bf 66} 123503 (2002) [arXiv:hep-th/0208156];
R. Kallosh and A. Linde,
JCAP {\bf 02} 02 (2003) [astro-ph/0301087];
U. Alam, V. Sahni and A.A. Starobinsky, JCAP {\bf 0304}, 002 (2003) [astro-ph/0302302].

\bibitem{ss02}
V. Sahni and Yu.V. Shtanov JCAP 0311,014, (2003) {\tt astro-ph/0202346}.

\bibitem{star17}
A. Shafieloo, D.K. Hazra, V. Sahni and A. A. Starobinsky,
{\em Metastable Dark Energy with Radioactive-like Decay}, arXiv:1610.05192.

\bibitem{sahni-sen17}
V. Sahni and A. A. Sen, 
Eur.Phys.J. {\bf C77}, 225 (2017) [arXiv:1510.09010]

\end{thebibliography}
\end{document}